\documentclass[english,apj]{emulateapj}
\usepackage{lscape}
\usepackage{apjfonts}
\makeatletter

\begin{document}

\newcommand{\irsfringe}{{\sc irsfringe}}
\newcommand{\osia}{{\sc osia}}
\newcommand{\RA}[4]{$\rm#1^h\,#2^m\,#3^s#4$}
\newcommand{\Dec}[4]{$#1\degr\,#2\arcmin\,#3\arcsec#4$}

\shortauthors{Fred Lahuis et al.}
\shorttitle{c2d IRS spectra of T~Tauri stars: gas-phase lines}

\title{c2d \textbf{\textit{Spitzer}} IRS Spectra of Disks around T~Tauri Stars. III. {[Ne\,II]}, [Fe\,I], and H$_2$ gas-phase lines}
\author{Fred~Lahuis\altaffilmark{1,2}, 
        Ewine~F.~van~Dishoeck\altaffilmark{1}, 
        Geoffrey~A.~Blake\altaffilmark{3}, 
        Neal~J.~Evans~II\altaffilmark{4}, 
        Jacqueline~E.~Kessler-Silacci\altaffilmark{4}, 
        Klaus~M.~Pontoppidan\altaffilmark{3} }

\altaffiltext{1}{Leiden Observatory, Leiden University, P.O. Box 9513, 2300 RA Leiden, The Netherlands}
\altaffiltext{2}{SRON Netherlands Institute for Space Research, P.O. Box 800, 9700 AV Groningen, The Netherlands}
\altaffiltext{3}{Division of Geological and Planetary Sciences 150-21, California Institute of Technology, Pasadena, CA 91125}
\altaffiltext{4}{The University of Texas at Austin, Department of Astronomy, 1 University Station C1400, Austin, Texas 78712--0259}
\email{F.Lahuis@sron.nl}

\begin{abstract}
We present a survey of mid-infrared gas-phase lines toward a sample of
76 circumstellar disks around low mass pre-main sequence stars from
the \textit{Spitzer} "Cores to Disks" legacy program. We report the
first detections of [Ne\,II] and [Fe\,I] toward classical T~Tauri stars in
\hbox{$\sim20$\,\%} respectively $\sim9$\,\% of our sources. The observed
[Ne\,II] line fluxes and upper limits are consistent with [Ne\,II]
excitation in an X-ray irradiated disk around stars with X-ray
luminosities $L_\mathrm{X}=10^{29}-10^{31}\,\mathrm{erg\,s^{-1}}$.
[Fe\,I] is detected at $\sim10^{-5}-10^{-4}\,L_\odot$, but no [S\,I] 
or [Fe\,II] is detected down to $\sim10^{-6}\,L_\odot$. The [Fe\,I] 
detections indicate the presence of gas-rich disks with masses of 
$\gtrsim0.1 M_\mathrm{J}$. No H$_2$\,0-0\,S(0) and S(1) disk emission
is detected, except for S(1) toward one source. These data give upper limits 
on the warm ($T\sim100-200$\,K) gas mass of a few Jovian masses, consistent
with recent T~Tauri disk models which include gas heating by stellar
radiation. Compact disk emission of hot ($T\gtrsim500$\,K) gas is 
observed through the H$_2$\,0-0\,S(2) and/or S(3)
lines toward $\sim8$\,\% of our sources. The line fluxes are, however,
higher by more than an order of magnitude than those predicted by
recent disk models, even when X-ray and excess UV radiation are
included. Similarly the \hbox{[Ne\,II]/H$_2$\,0-0\,S(2)} ratios for these
sources are lower than predicted, consistent with the presence of an
additional hot molecular gas component not included in current disk
models. Oblique shocks of stellar winds interacting with the disk
can explain many aspects of the hot gas emission, but are
inconsistent with the non-detection of [S\,I] and [Fe\,II] lines.

\end{abstract}

\setcounter{footnote}{4}

\clearpage

\section{Introduction}
\label{sec:introduction}

Circumstellar disks are a natural and important phenomenon in low-mass
star formation. The ability to trace the evolution of the disk dust
and gas content is crucial to understanding their chemistry and
physics and the formation of planets. Observed disks reveal a large
source to source variation and suggest a complex evolution from young
gas-rich disks to tenuous debris disks.  Disk geometries are observed
to range from almost flat to strongly flaring disks
\citep[][]{dullemond04}, and disks with large inner cavities are found
\citep[][]{calvet02,bouwman03,forrest04,brown07}. Observations of
silicates and spectral energy distributions (SEDs) present evidence
for grain growth and settling of large grains to the disk midplane
\citep[][]{vanboekel03,dalessio06,kessler06}. Different degrees of
grain heating and radial mixing in the disks \citep[][]{vanboekel05},
and varying PAH abundances affect the heating of the upper layers of
the disk \citep[][]{habart06,geers06}. Most of these disk properties are
derived from infrared solid-state features and from near-infrared to
millimeter continuum observations and associated SED modeling.
However, such data give little information about the gas in the disk.

Gas plays an important role in the structure and evolution of disks,
including the temperature and density gradients, chemistry, dust
dynamics, and eventually the formation of gas-rich and gas-poor
planets \citep{gorti04}. Observational diagnostics of the physical
conditions of the gas, in particular its mass and temperature, are
therefore highly relevant to studies of disk evolution and planet
formation.  Central questions are how the gas in the disk is
dissipated, what drives the disk heating and gas clearing, and what
the associated timescales are.  The small number of observed
transitional objects between the classical T~Tauri phase (large
H$\alpha$ equivalent width, thought to be accreting) or the weak-line
T~Tauri phase (small H$\alpha$, non accreting), both with massive optically
thick disks, and the more evolved phase with optically thin
or no disks, suggests that disk clearing timescales are short 
(few$\times 10^5$\,yr) compared to the disk lifetime of a few Myr 
\citep[][]{cieza07,haisch01}. In addition, combined near-infrared
(IR), mid-IR, and millimeter observations imply that the disk clearing happens
nearly simultaneously across the disk \citep[see
e.g.][]{kenyon95,hartmann05,takeuchi05}.
\citet[][]{alexander06} present a new evolutionary model combining
viscous evolution with photoevaporation of the disk to address these
issues.  In this mechanism, the disk is cleared through an evaporative
flow originating from the disk surface layers as a result of UV and/or
X-ray heating \citep[see
e.g.][]{hollenbach00,clarke01,kamp04,dullemond07}.  Information about
the temperature and mass of the warm gas and constraints on the
details of the heating processes across the disk are therefore vital
for advancing the current models.

Models of gas heating include UV and X-ray radiation from the star
itself as well as possible excess radiation due to accretion
\citep[e.g.][]{jonkheid04,gorti04,kamp04,nomura05,jonkheid07}. The
resulting gas temperatures in the surface layers out to large radii
are significantly higher than those of the dust as a result of the
photoelectric effect on small grains and PAHs. Gas temperatures may
reach values of up to a few thousand~K.  Once grains have grown to
$\mu$m sizes or larger, however, the gas temperature is significantly
decreased unless PAHs are still present (Jonkheid et al.\ 2004, 2006,
2007). \citet{nomura05} include explicitly the excitation of H$_2$ by
UV and collisions for a disk around a T~Tauri star with and without
excess UV and make predictions for line intensities which can be
tested against observations.  \citet[][]{glassgold07} have studied the
case of X-ray heating and have shown that the ionized neon
fine-structure line emission may provide unique tracers of X-ray
heating in the disk surface since neon cannot be photoionized by
radiation with energies below 13.6\,eV.

Observational studies of the gas and its temperature have mostly
focused on the hot inner and the cold outer regions of disks.
High-resolution CO \textit{v}=1--0 vibration-rotation lines at 4.7
$\mu$m \citep{najita03,brittain03,blake04} and
\hbox{H$_2$\,$2.1\mu$m\,1--0\,S(1)} \citep{bary03} show gas with temperatures
$\sim1000-3000$\,K in the surface layers out to $\sim1\,$AU.  H$_2$O
emission from SVS\,13 \citep[][]{carr04} and molecular absorption of
C$_2$H$_2$, HCN, and CO$_2$ in the disk of IRS\,46
\citep[][]{lahuis06a} also indicate hot temperatures in the inner few
AU of several hundred K.  In contrast, millimeter CO surveys probe the
cold gas throughout the outer disk where the dust is optically thin
\citep[e.g.][]{koerner95,duvert00,thi01,dutrey03,dent05}.  Its use as
a gas mass tracer is however limited as a result of both strong
photodissociation at low extinction and freeze-out in the disk
interior.  Gas temperature determinations range from $<$20~K near the
midplane \citep{dartois03,pietu07} to 40 K or higher in the
intermediate and surface layers \citep{zadelhoff01,qi06}.

%

The mid-IR H$_2$ and atomic fine structure lines are best suited as
direct tracers of the warm ($\sim$100 K) gas in the intermediate zones
of disks at radii of a few AU out to several tens of AU, i.e., the
planet-forming zones of disks. The Infrared Space Observatory (ISO)
provided the first opportunity to probe this warm gas in disks around
Herbig Ae and T~Tauri stars.  \citet{thi01} suggested that
large amounts ($\sim0.01-100$\,M$_J$) of gas could reside in disks 
around young T~Tauri stars, but this has not been confirmed by subsequent
ground-based observations \citep[][]{richter02,sheret03,sako05}.

The sensitive InfraRed Spectrograph (IRS) \citep[][]{houck04} on
board the \textit{Spitzer} Space Telescope \citep[][]{werner04} brings
the detection of these lines within reach for young solar mass stars
in nearby star forming regions.  The combination of high
sensitivity, moderate spectral resolution
$R=\lambda/\Delta\lambda=600$, and modest spatial resolution
makes \textit{Spitzer} well suited for the direct study of the gas in and
around low-mass young stars in nearby ($\lesssim300$\,pc) clouds
through the mid-IR lines of various species.

We present here an overview of gas-phase lines detected in disks
observed in the \textit{Spitzer} legacy program ``From Molecular Cores
to Planet Forming Disks'' (``Cores to Disks'' or c2d)
\citep[][]{evans03}, which has collected a large sample of IRS spectra
toward sources in the nearby Chamaeleon, Lupus, Perseus, Ophiuchus,
and Serpens star-forming regions.  High-S/N 5-38 $\mu$m spectra have
been obtained for 226 sources at all phases of star and planet
formation up to ages of $\sim$5 Myr.  From this sample, 76 disk
sources, identified by showing either the 10 or 20 $\mu$m silicate
bands in emission, have been selected. In Sections
\ref{sec:observations} and \ref{sec:data-reduction} the source
selection and data reduction are explained.  In Section
\ref{sec:results} the observed atomic fine-structure and H$_2$
emission lines and the derived parameters are presented. In Section
\ref{sec:discussion} the results are reviewed in the context of
currently available disk models. This paper forms a complement to the
searches for the mid-infrared lines of H$_2$ and other species toward
more evolved disks studied in other \textit{Spitzer} programs
\citep[e.g.][]{hollenbach05,pascucci06,pascucci07}. 

\begin{figure*}[b]
\begin{center}
\parbox{0.37\textwidth}{\includegraphics[width=0.37\textwidth]{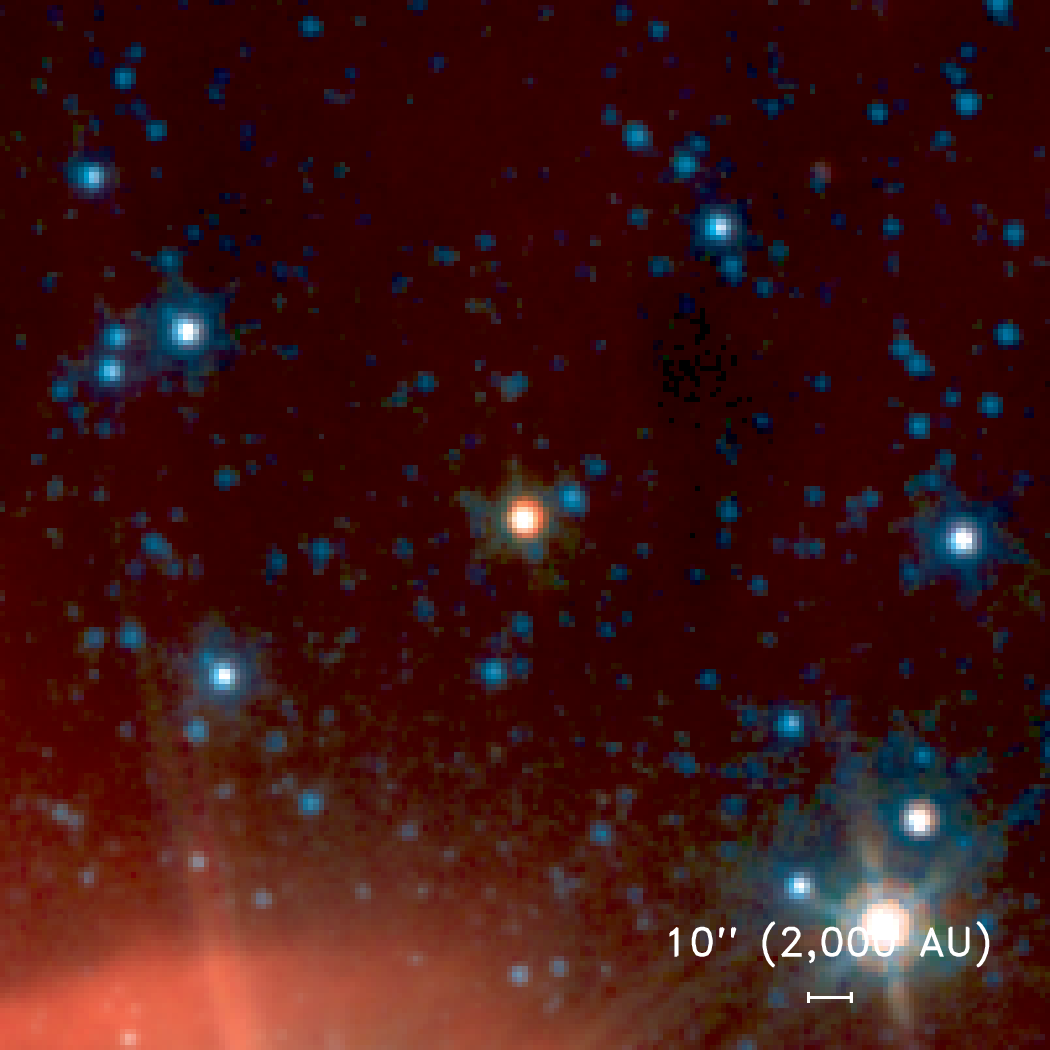}}%
\parbox{0.37\textwidth}{\includegraphics[width=0.37\textwidth]{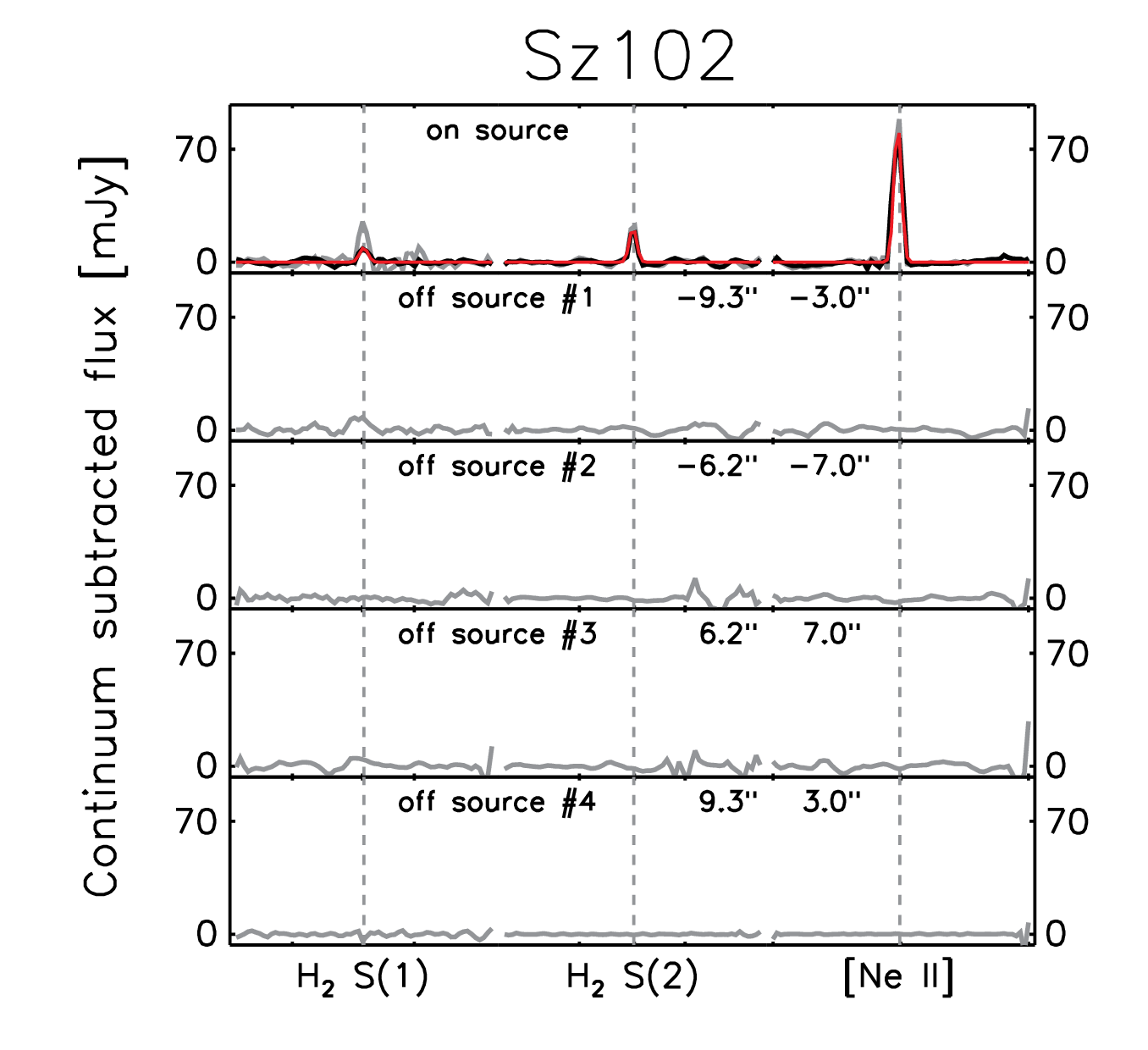}}\\
\parbox{0.37\textwidth}{\includegraphics[width=0.37\textwidth]{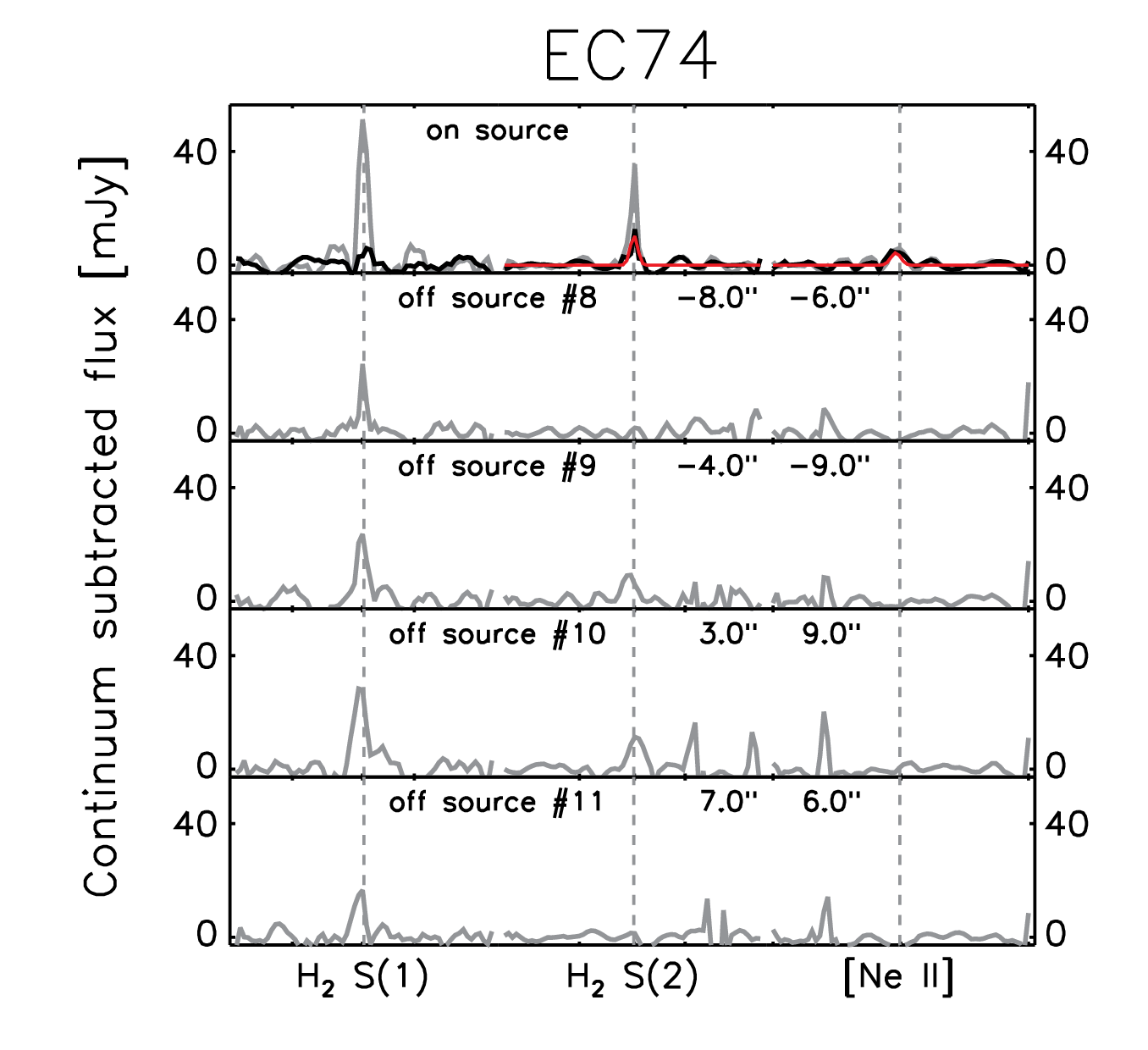}}%
\parbox{0.37\textwidth}{\includegraphics[width=0.37\textwidth]{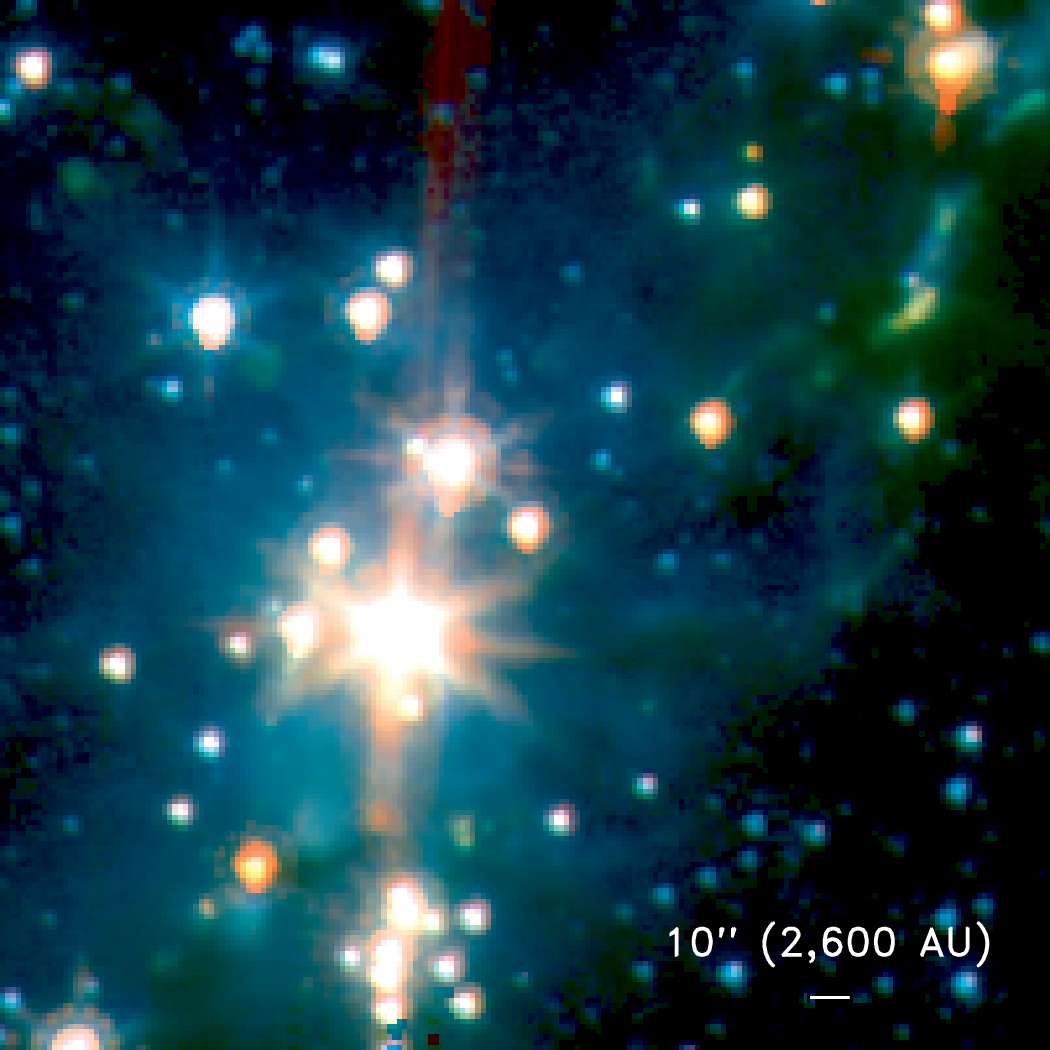}}\\
\parbox{0.37\textwidth}{\includegraphics[width=0.37\textwidth]{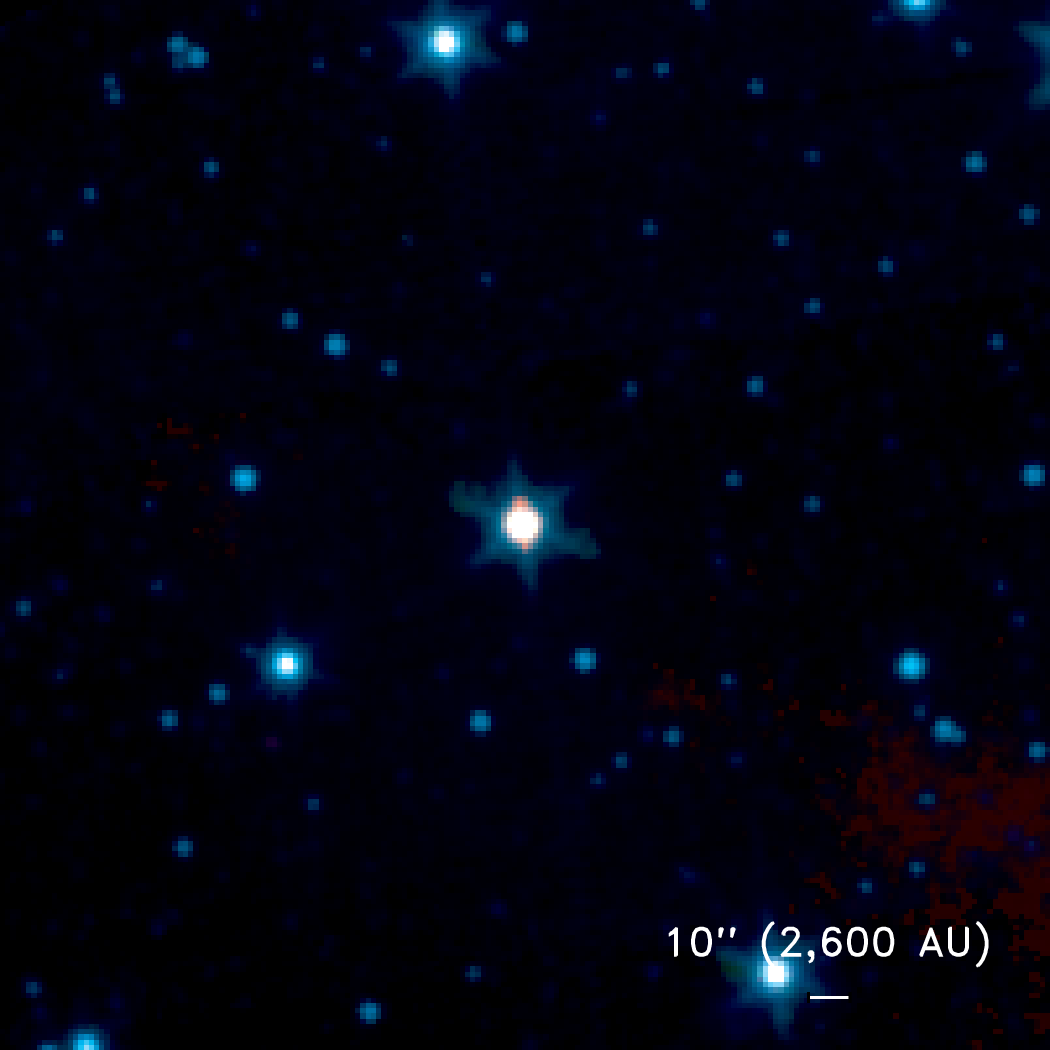}}%
\parbox{0.37\textwidth}{\includegraphics[width=0.37\textwidth]{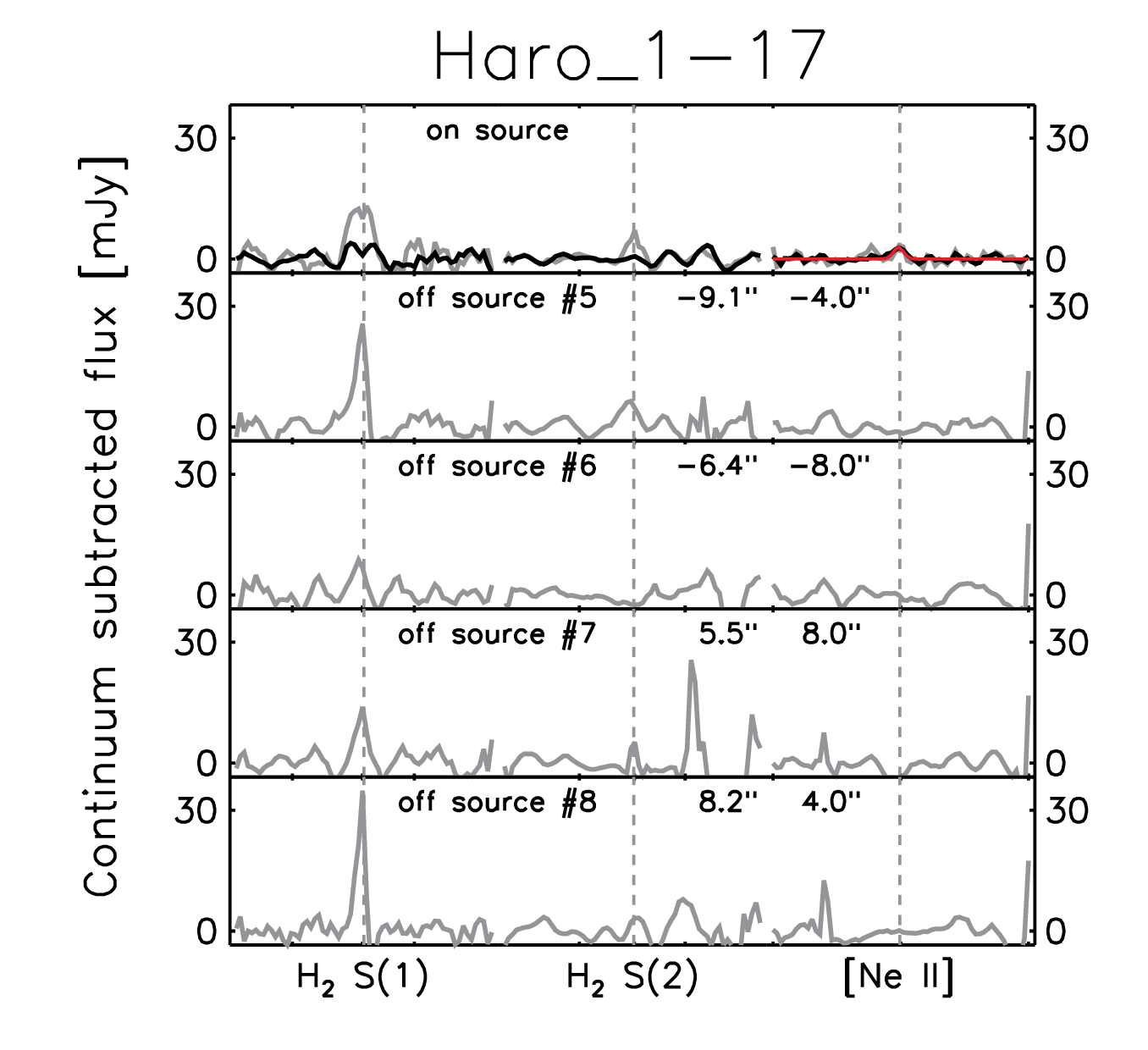}}
\vspace{\baselineskip}
\caption{\small\label{fig:off-source}%
Observations of H$_2$\,0-0\,S(1), S(2), and [Ne\,II] emission observed on and
off source toward Sz\,102, EC\,74, and Haro\,1-17. The on source observations
were observed in the c2d first-look program, the follow up off source
observations in the second look program (see Sec. \ref{sec:observations}).
The images show \textit{Spitzer} composites of IRAC1 (blue), IRAC2 (green),
and IRAC4 (red)  with the sources in the middle.
The gray lines in the spectra show the total
(compact source plus extended component) observed emission.
The black lines the source emission after correction for the
extended component.
The red lines show Gaussian fits to the compact source emission
if observed at $3\sigma$ or more.
Although the S/N in the off source positions is often poor, it illustrates
the problem of extended and non-uniform H$_2$ emission. The [Ne\,II]
is almost always limited to the source itself.
}
\end{center}
\end{figure*}

\section{Observations}
\label{sec:observations}

The data presented in this study were selected from the sample of IRS
spectra observed within the \textit{Spitzer} c2d legacy program.
The c2d IRS program consists of two programs of comparable size,
referred to as the first- and second-look programs.  The first-look
program ({PID\,\#172}) was restricted primarily to known low-mass
young stars, embedded YSOs and pre-main-sequence stars with disks with
masses $M<2$~$M_{\sun}$ and ages $\lesssim5$\,Myr, and a sample
of background stars. A few Herbig Ae stars are included as well. The
c2d source selection criteria were defined to be complementary to
those of the \textit{Spitzer} legacy program ``The Formation and
Evolution of Planetary Systems'' \citep[FEPS,][]{meyer02}.
The second-look program ({PID\,\#179}) was, for the most part, devoted
to IRS follow-up spectroscopy of sources discovered in the IRAC and MIPS
mapping surveys, including a newly discovered cluster of young stars
in Serpens \citep[][]{harvey06}.
For all first-look observations, the integration times for the short-high (SH)
and long-high (LH) modules ($R=600$, 10--37 $\mu$m) were chosen such that
theoretical signal to noise ratios (S/Ns) of at least 100 and 50 on the
continuum were obtained for sources brighter and fainter than 500 mJy,
respectively. Deeper integrations were not feasible within the
c2d program. Spectra taken using the short-low (SL) or long-low (LL) modules
($R=60-120$, 5-14 $\mu$m and 14-38 $\mu$m respectively) always reach
theoretical S/N ratios greater than 100.
For the second-look IRS targets similar S/N limits were obtained wherever
possible. However, since the second-look contained a number of very
weak sources (down to a few mJy) this was not always achieved.

\subsection{Source selection}
\label{sec:source-selection}

The sources presented in this paper were all selected to show either
of the $10\,\mu$m or $20\,\mu$m silicate bands in emission.  A total
of 76 sources were chosen; see \citet[][]{kessler06} for the 47 
first-look disk sources with silicate emission.  This selection
excludes most edge-on disk sources ($i\gtrsim65$ degrees) such as
CRBR\,2422.8-3423 \citep[][]{pontoppidan05a}, IRS\,46
\citep[][]{lahuis06a} and the `Flying Saucer'
\citep[][]{pontoppidan07c}, with the exception of the high inclination
sources EC82 \citep[][]{pontoppidan05b} and VV\,Ser which are included
in this paper.
Gas-phase lines toward edge-on disk sources will be discussed in a
separate paper together with the embedded class 0 and I sources
\citep[][]{lahuis07}.
The selected sources are listed in Table \ref{disktab:sourcelist} which gives
the basic observing and source parameters, e.g. the adopted distances.

\begin{figure*}
\begin{center}
{\hfill%
\parbox{0.05\textwidth}{\hfill}%
\parbox{0.48\textwidth}{\includegraphics[angle=90,width=0.48\textwidth]{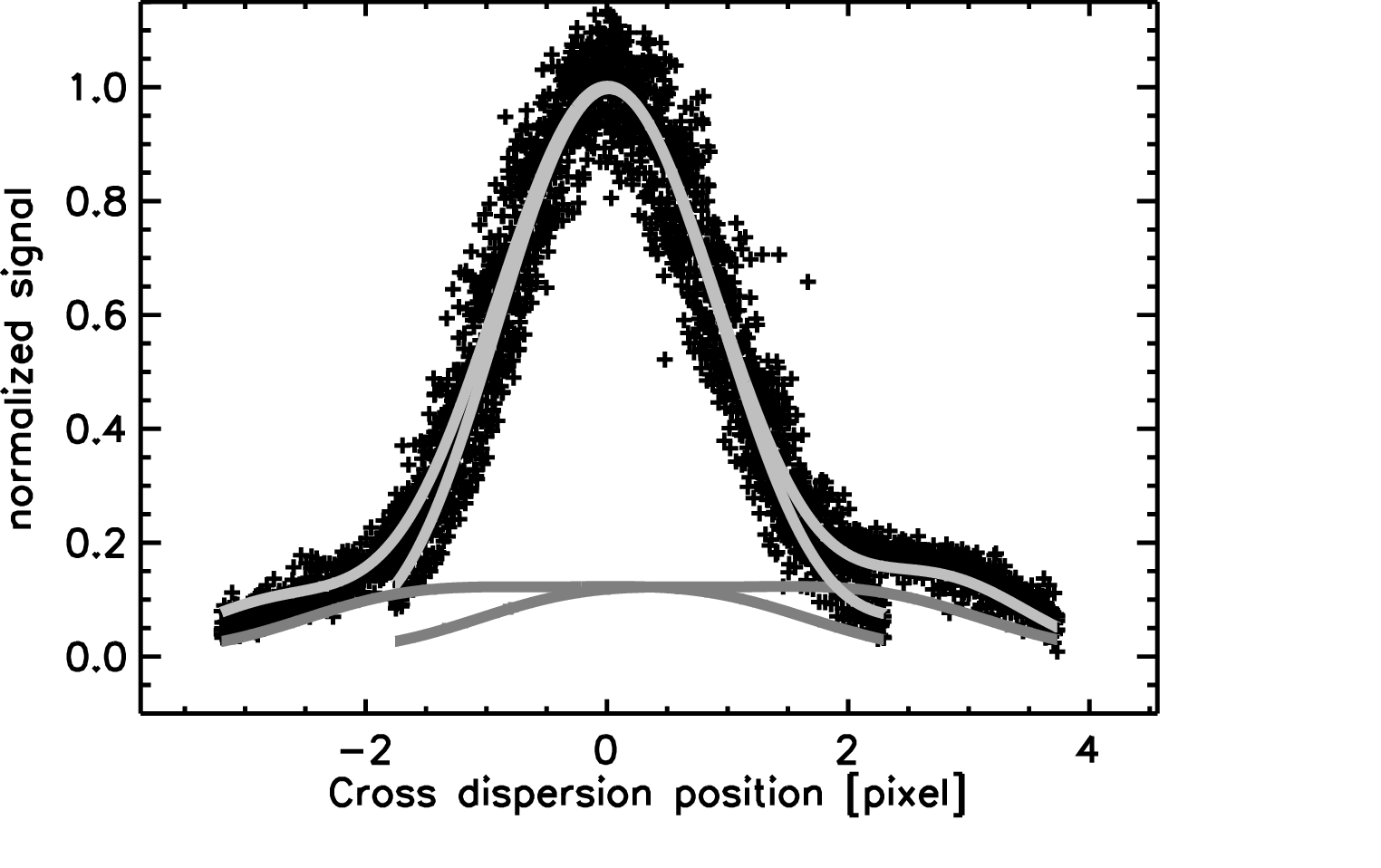}}%
\parbox{0.4\textwidth}{\includegraphics[width=0.4\textwidth]{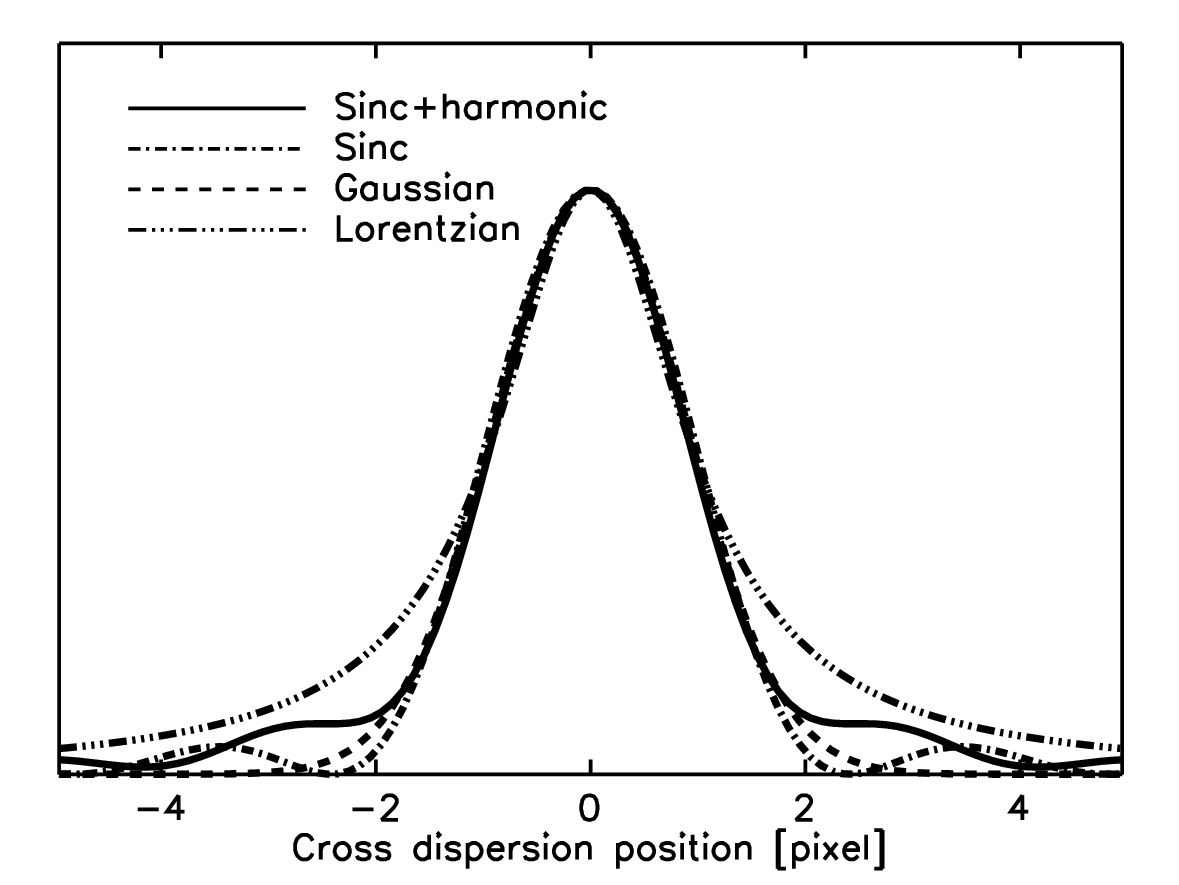}\newline}%
\parbox{0.1\textwidth}{\hfill}%
\hfill}
\caption{\small\label{fig:xdisp-profile}%
{\small
Illustration of the \textit{Spitzer} IRS cross dispersion profile used
in the optimal extraction (see Sec. \ref{sec:optimal-extraction}).
The left plot shows a fit to the IRS SH order 11 data of GW\,Lup,
a source with a moderate but clear sky component in the IRS spectra.
The rsc data (bcd data before flatfielding) of both dither position
(black plusses) is shown, normalized, collapsed along the dispersion direction,
and corrected for the cross dispersion dither offsets.
Overplotted is the combined fit of the source profile plus the
extended emission in gray and the extended emission in dark gray.
The shape of the extended emission reflects the IRS
flatfield of the (for this source) assumed uniform extended emission.
The right plot shows a comparison of an IRS PSF profile (Sinc plus harmonics)
compared to the profiles of an undistorted Sinc, a Gaussian, and a Lorentzian
profile with the same FWHM. Note the significant variation in the strength
and shape of the profile wings.
The correct characterization of both the width and the wings of the profile
for all IRS orders is essential for extracting the proper source and
sky spectra.
}
}
\end{center}
\end{figure*}

\subsection{SH mini maps}
\label{sec:minimaps}
In an early phase of the c2d project molecular hydrogen and [Ne\,II]
lines were detected toward some of the c2d sources. As part of the
second-look program, follow-up mini-maps were taken using the SH
module to check for extended emission at offsets positions of
$\sim$10-15$''$ with respect to the sources. 
Five maps were defined to include off source
observations around eight sources. Three of these are disk sources
included in our sample, Sz\,102 (Krautter's star), Haro\,1-17, and
EC74.  Figure \ref{fig:off-source} shows the observed
H$_2$\,0-0\,S(1), S(2), and [Ne\,II] emission from the first-look on
source observations together with the off source observations from the
second-look mini maps.  The maps show that most of the H$_2$ emission
is extended, especially for the S(1) line.  However, fine-structure
lines are usually seen to be limited to the source itself.
Subsequent results using the c2d optimal extraction procedure (see
Section \ref{sec:optimal-extraction}) confirm the conclusions drawn
from the analysis of the mini-maps.

The SH maps are not complete since the time allocated for c2d
spectroscopy did not allow to observe fully sampled maps.
The prime purpose of the maps is
to confirm the presence or absence of extended emission.
More extended mapping observations will be required to study the
detailed spatial distribution and extent of the large scale
emission component.

\section{Data reduction}
\label{sec:data-reduction}
The c2d reduction pipeline \citep[][]{lahuis06b} was used to reduce
the IRS data, starting from the S13 and S14 archive data.
The same c2d pipeline products as included in the final c2d Legacy 
data delivery%
\footnote{The c2d legacy data are accessible at\\
{\url{http://ssc.spitzer.caltech.edu/legacy/c2dhistory.html}.}}  were
used for the spectral line analysis (see Sec. \ref{sec:analysis}).
Most of the analysis focused on the SH and LH data, since the SL and
LL data are generally limited by the low line/continuum ratio.  The SL
data were included and used to search for higher-excitation H$_2$\,0-0
lines, in particular the S(3) transition.

\subsection{Separating disk and cloud emission -- optimal extraction}
\label{sec:optimal-extraction}

A major concern when studying emission lines from young stellar
objects (YSOs) is the possible contribution of extended (envelope or
local cloud) emission in the sometimes complex star forming regions.
The spatial distribution of the emission, both in the continuum and in
spectral lines, often prohibits the use of `sky' observations alone to
correct for extended emission components. For this reason the c2d team
has developed an optimal extraction algorithm for IRS pointed
observations.

The \textit{Spitzer} diffraction limited beam is $\sim4-5$ arcsec for the SH
module (10-19.5\,\micron) and $\sim7-10$ arcsec for the LH module
(19-37\,\micron).  At a distance of 100 parsec this corresponds to
physical sizes of $\sim$400-500 AU and $\sim$700-1000 AU,
respectively.  The clouds observed in the c2d program are located at
distances ranging from 125 parsec (Ophiuchus) to 260 parsec (Serpens)
increasing the physical area observed.  The full IRS aperture in the
SH and LH spatial direction is $\sim2.5-3$ times larger than the beam
size. At the observed cloud distances this means that the aperture
probes physical scales of several thousand AU.  This makes it ideally
suited for detecting cold or shocked H$_2$ emission from the
extended (remnant) envelope, outflows, or the diffuse local cloud
emission. For the disk sources studied in this work however
the local cloud emission will potentially confuse the compact disk
emission.  Distinguishing between compact (disk) and extended
(remnant envelope, outflow, or diffuse cloud) emission is therefore
of vital importance for studying the emission lines originating
in the circumstellar disks. The optimal PSF
extraction developed by the \textit{Spitzer} c2d legacy team
\citep[][]{lahuis06b} allows separation of the two components for all
sources. The mini-maps (see Section \ref{sec:minimaps}) observed
around selected sources confirm the results of the optimal
extraction.  See Figure \ref{fig:off-source} for an example of extended
H$_2$ emission but compact [Ne\,II] emission. Other examples include
Figures 3 and 4 in \citet[][]{geers06} for separating extended cloud
and compact disk PAH emission.

The optimal extraction uses an analytical cross-dispersion point
spread function (PSF) for the source profile plus an extended emission
component, to fit the observed crosstalk or straylight corrected 
echelle (SH and LH) and longslit (SL and LL) images.
The PSF is described by a sinc function with a harmonic distortion
component which results in a singnificant broadening of the profile
wings (see Fig. \ref{fig:xdisp-profile}). 
The wavelength dependence of the PSF parameters, the order trace, 
the width, and the harmonic distortion, are characterized using 
a suite of high S/N calibrator stars. For the extended emission 
component the flatfield cross-dispersion profile is used.
The flux calibration derived from the calibrator stars using Cohen 
templates and MARCS models \citep[][]{decin04} provided through the 
{\it Spitzer} Science Center. \citet{lahuis06b} give more details 
about the characterization and calibration of the c2d optimal
extraction.
The optimal extraction returns the total flux (the source flux plus 
extended emission in the IRS beam) plus an estimate of the extended emission
component. Error estimates are derived for both the total emission and
the extended emission component. The S/N of the extended emission
component can vary significantly depending on the quality of the raw
image data and on deviations of the extended emission from the
assumed uniformity across the IRS slit. Therefore, care has to be
taken when subtracting the extended emission from the total flux
signal to retrieve the compact source emission. In some cases, a fit
to the extended continuum and line emission is used to avoid adding in
surplus noise from the extended emission component.
The uncertainty on the fit to the extended emission is propagated
into the error of the compact source signal.

\subsection{1-D spectra}
\label{sec:1d-spectra}

After extraction, the 1-D spectra are corrected for instrumental
fringe residuals \citep{lahuis03irs}, order matching is applied, and a
pointing flux-loss correction is performed to the compact source
component.  Pointing offsets up to a few arcsec can have a noticeable
impact on the derived fluxes of lines observed with the SH and SL
modules, e.g. H$_2$\,0-0\,S(1), S(2), [Ne\,II], and [Ne\,III]. For
example, dispersion offsets within the nominal $3\sigma$ pointing
uncertainty of \textit{Spitzer} ($\sim1\arcsec$ for medium accuracy
peakup) can lead to SL and SH flux losses up to $\sim10$\% depending
on wavelength. For all targets, a combination of either the SH, LH,
and SL, SH and LH, or SH and SL modules is available.  This allows
correction of the pointing related flux losses with an accuracy given
by the S/N of the data in the module overlap areas.  A detailed
description of the c2d pipeline (including extraction, defringing,
pointing flux loss correction) and the c2d legacy products is given in
the ``c2d Spectroscopy Explanatory Supplement'' \citep[][]{lahuis06b}.

\begin{figure}
\begin{center}
\includegraphics[width=0.9\columnwidth]{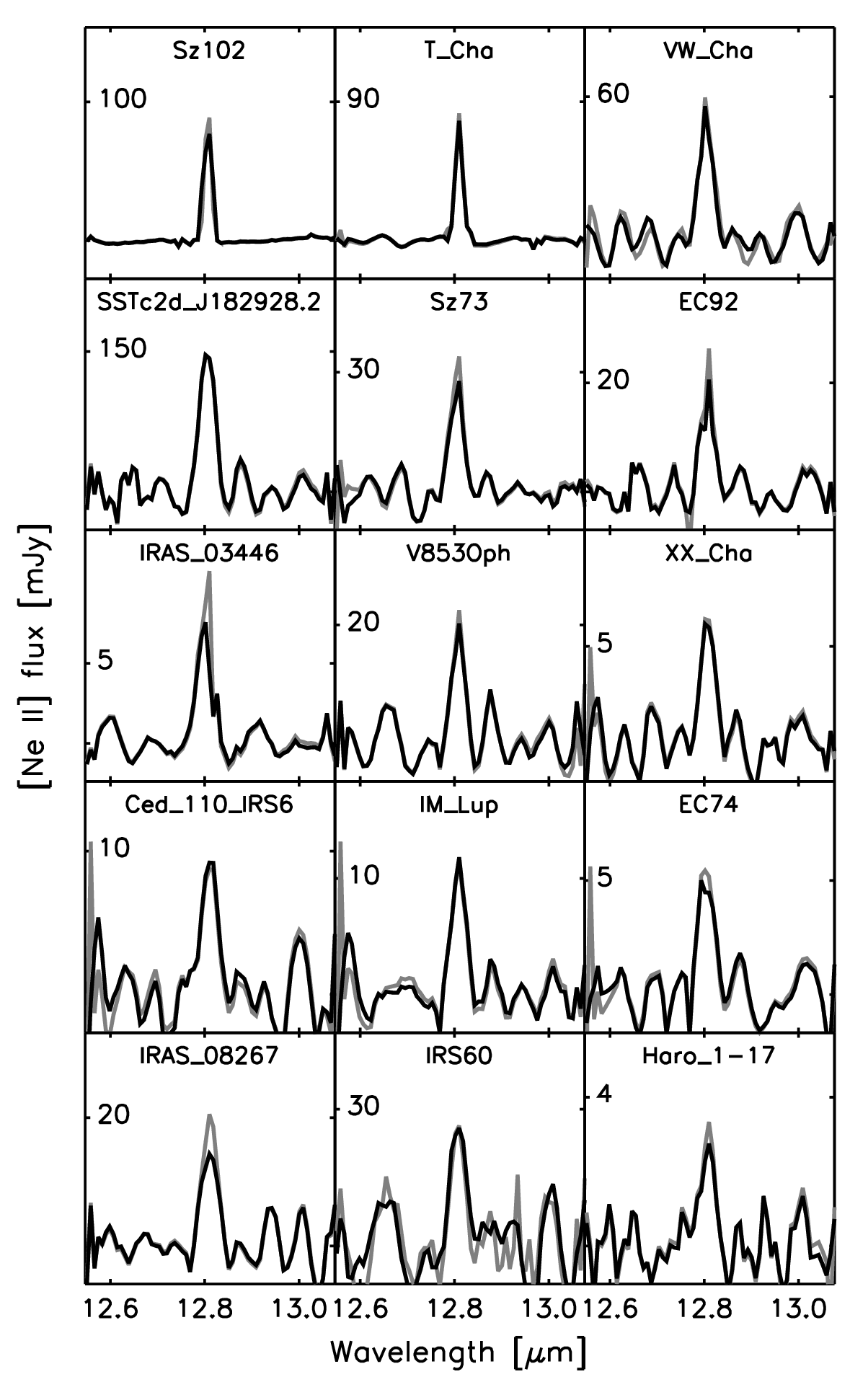}\\
\includegraphics[width=0.9\columnwidth]{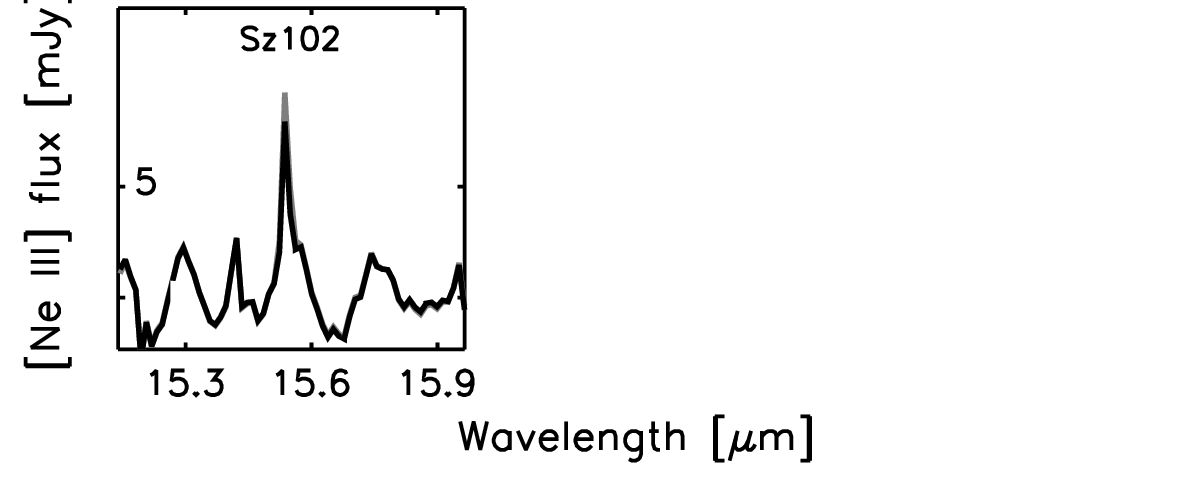}
\caption{\small\label{fig:NeII_lines}%
Detections of neon lines at the $3\sigma$ level or more toward
the c2d sample of T~Tauri stars with disks.
Of all the H$_2$  and atomic fine structure lines [Ne\,II] is most
convincingly detected toward $\sim20$\% of the sources.
Plotted in gray is the total observed emission (compact source +
extended component) and in
black the emission after correction of the estimated sky component.
None of the sources show a significant extended component.
}
\end{center}
\end{figure}

\subsection{Spectral Analysis}
\label{sec:analysis}

The SH and LH modules of the IRS instrument cover the positions of the
three lowest pure rotational lines of H$_2$ and emission of [Ne\,II]
(12.8 $\mu$m), [Ne\,III] (15.55 $\mu$m), [Fe\,I] (24 $\mu$m), 
[Fe\,II] (17.9 and 26.0 $\mu$m),
[S\,I] (25.25 $\mu$m), [S\,III] (18.7 $\mu$m), and [Si\,II](34.8 $\mu$m) at a
resolving power of $R=\lambda/\Delta\lambda = 600$.  Line fitting and flux
integration is done using routines from \osia\footnote{\osia\ is a
joint development of the ISO--SWS consortium.  Contributing institutes
are SRON, MPE, KUL and the ESA Astrophysics Division.
{\url{http://sws.ster.kuleuven.ac.be/osia/}}}.

As discussed in Section \ref{sec:optimal-extraction}, the extended emission
component, both in the continuum and the spectral line, is subtracted
from the spectrum prior to line fitting. Uncertainty estimates, as
listed in Table \ref{disktab:linefluxes}, are derived from the residuals
after line fitting, or, in the absence of a spectral line, using the
line width derived from the instrumental resolution. The uncertainty
derived from the extended emission is added into the uncertainty
estimate of the source component. As a result, the 1-$\sigma$
uncertainty estimates can vary widely for sources with a similar
continuum flux and integration time. This may for example be the
result of the presence of artifacts resulting from hot pixels or
small variations in the extended emission which are not accounted 
for in the spectral extraction which assumes a constant extended 
emission component.

Typical mean 3$\sigma$ uncertainties prior to subtraction of the
extended component for the high resolution modules range from $\sim
1\times10^{-16}-2\times10^{-15}\ \mathrm{erg\,cm^{-2}\,s^{-1}}$ with
positive and negative extremes of $\sim5\times10^{-17}\
\mathrm{erg\,cm^{-2}\,s^{-1}}$ and $\sim1\times10^{-14}\
\mathrm{erg\,cm^{-2}\,s^{-1}}$.  The uncertainties are comparable to
those from \citet[][]{pascucci06} for FEPS observations using
on source integration times similar to those used for the c2d sample.

\begin{figure}
\begin{center}
\includegraphics[width=\columnwidth]{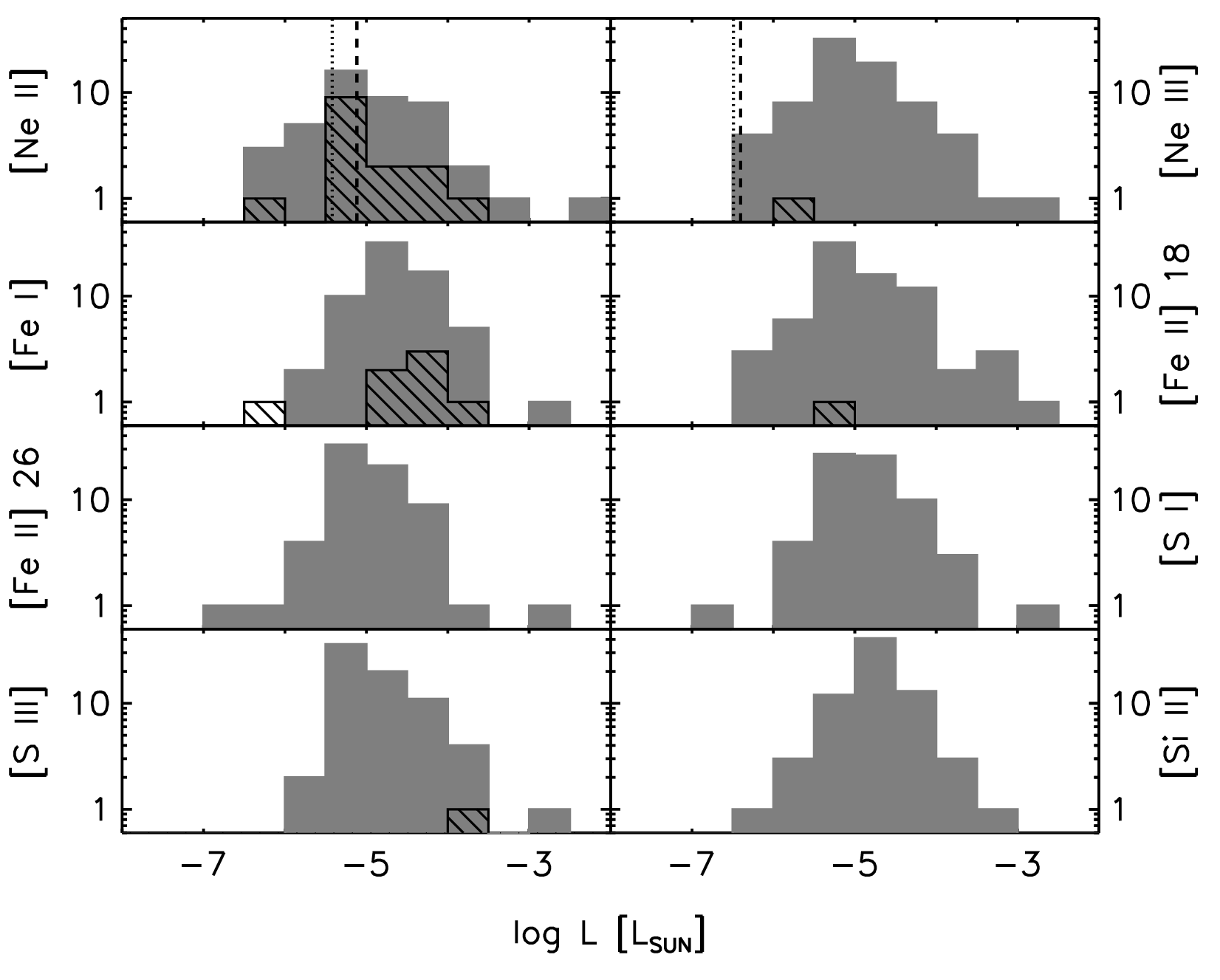}
\caption{\small\label{fig:model_lineflux_atomic}%
Observed line strengths (hatched bars) and upper limits (solid gray bars)
of the major atomic lines. Included with vertical lines are model
predicted line strengths for [Ne\,II] and [Ne\,III] by \citet[][]{glassgold07}.
The model incorporates UV and X-ray heating of the circumstellar disk.
The two lines represent two extreme thermal models, X-ray heating
dominant (dotted line) or mechanical heating dominant (dashed line).
}
\end{center}
\end{figure}

\section{Results}
\label{sec:results}

\subsection{Atomic fine-structure lines: Neon}
\label{sec:neII}

Of all the atomic fine structure lines covered by the SH and LH
modules, the [Ne\,II] $12.8\mu$m transition is most convincingly
detected and shows the strongest source emission. The higher
excitation [Ne\,III] $15.5\mu$m line is tentatively detected toward
Sz\,102 but not toward any of the others sources in our sample. Figure
\ref{fig:NeII_lines} shows all [Ne\,II] and [Ne\,III] lines detected
at $3\sigma$ or more; see also Fig.~1 of \citet[][]{geers06} for the
[Ne\,II] 12.8 $\mu$m line toward T Cha.  Plotted in Figure
\ref{fig:NeII_lines} is the total continuum-subtracted observed
[Ne\,II] emission in gray and the compact source emission after
correction for extended line emission in black. Taken together,
[Ne\,II] emission is observed in the spectra of 15 T~Tauri sources
($\sim20$\,\% of the sample). These
are the first reported detections of [Ne\,II] toward disks around
classical T~Tauri stars. The optimal extraction method, together with
the limited mini-maps (Sect. \ref{sec:minimaps} and Fig.\
\ref{fig:off-source}), show that the emission is indeed associated
with the source itself.

The observed line fluxes and upper limits of [Ne\,II] and [Ne\,III]
are listed in columns 8 and 9 of Table \ref{disktab:linefluxes}.  When
line fluxes are compared (observed or with models) the line strength
is converted to solar luminosities since the sample is observed toward
sources from multiple clouds and compared to model predictions using
different assumed distances.  Figure \ref{fig:model_lineflux_atomic}
shows the distribution of the observed line strengths (hatched bars) and
upper limits (solid gray bars).

\begin{figure}
\begin{center}
\includegraphics[width=0.9\columnwidth]{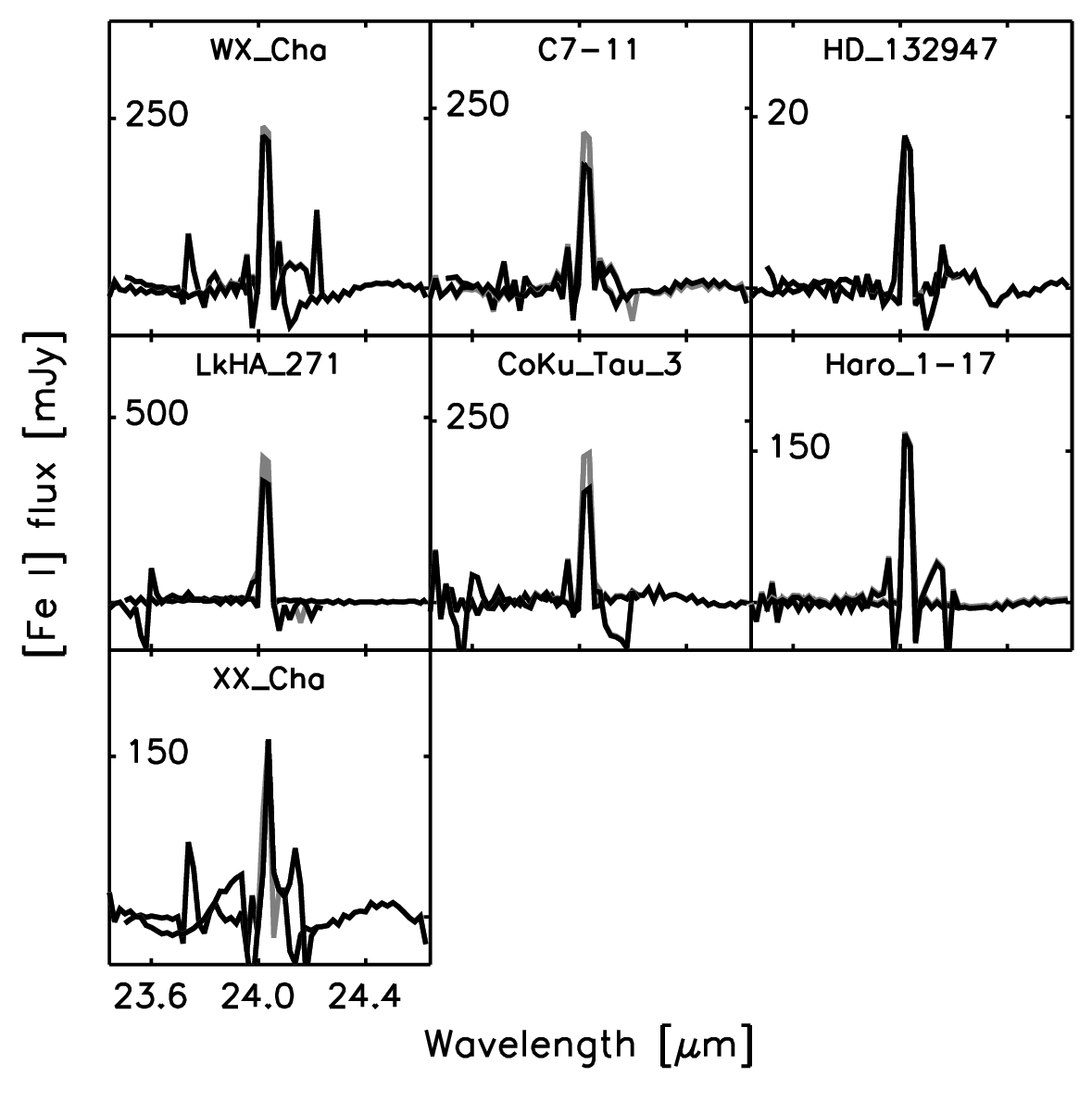}\\
\includegraphics[width=0.9\columnwidth]{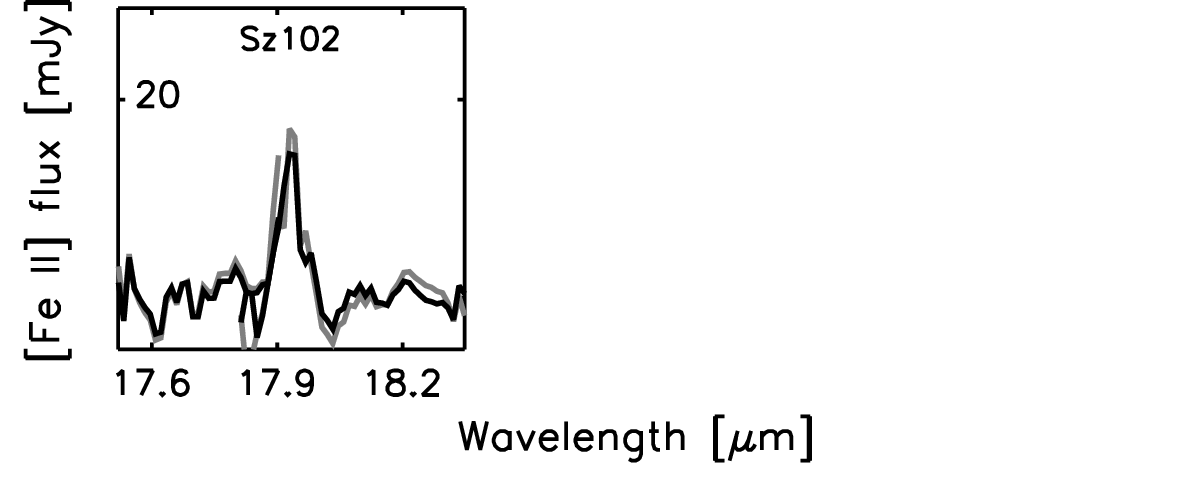}\\
\includegraphics[width=0.9\columnwidth]{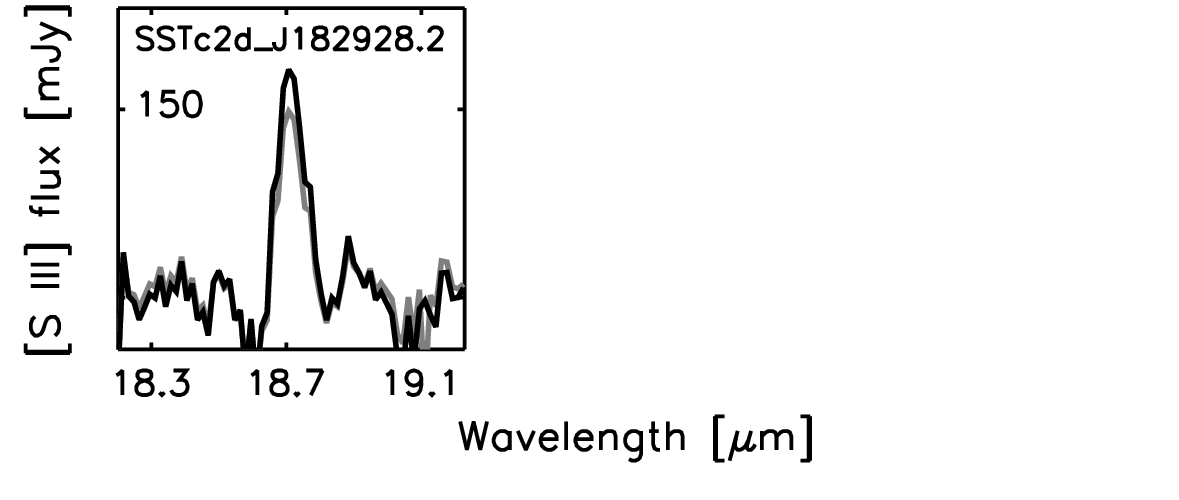}
\caption{\small\label{fig:Atomic_lines}%
Detections of [Fe\,I], [Fe\,II], and [S\,III] toward the c2d sample of
T~Tauri stars with disks. Plotted in gray is the total observed emission
(compact source + extended emission) and in black the emission after
correction of the estimated extended component. No significant extended
line emission is observed.
}
\end{center}
\end{figure}

\subsection{Atomic fine-structure lines: Other species}
\label{sec:other-species}

[Fe\,I] at $24\mu$m is the only other species besides [Ne II] with
clear detections toward seven sources ($\sim9$\,\% of the sample). Of
the other atomic lines there is one detection of [Fe\,II] at $18\mu$m
and one of [S\,III] at 18.7 $\mu$m, in different sources. [S\,I] at 25
$\mu$m, [Fe\,II] at $26\mu$m, and [Si\,II] at 34 $\mu$m are not
detected.  The derived line fluxes and upper limits are listed in
columns $9-14$ in Table \ref{disktab:linefluxes}.  The detected lines are
plotted in Figure \ref{fig:Atomic_lines} whereas Figure
\ref{fig:model_lineflux_atomic} shows the distribution of the observed
line strengths.

\begin{figure}
\begin{center}
\includegraphics[width=\columnwidth]{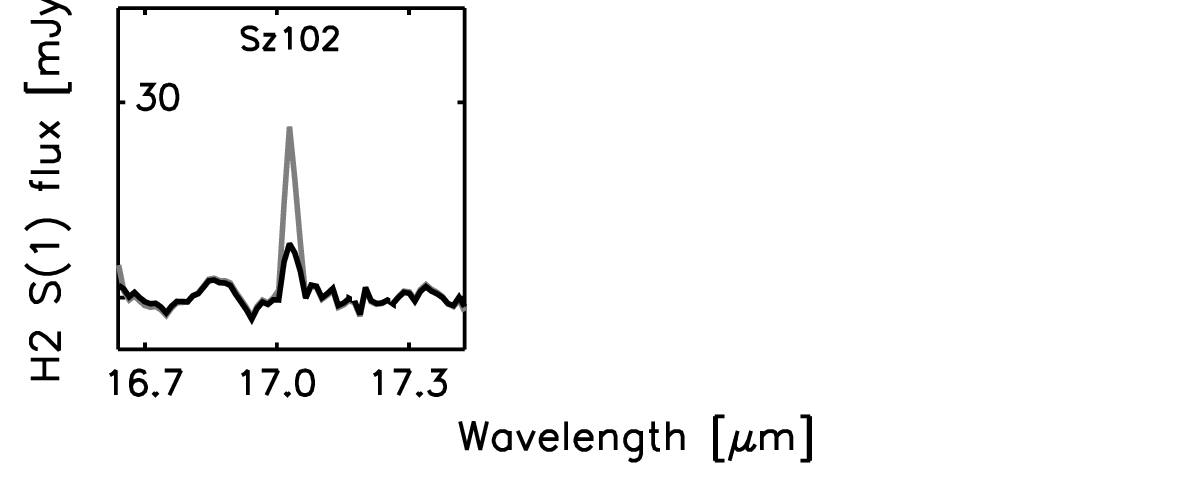}\\
\includegraphics[width=\columnwidth]{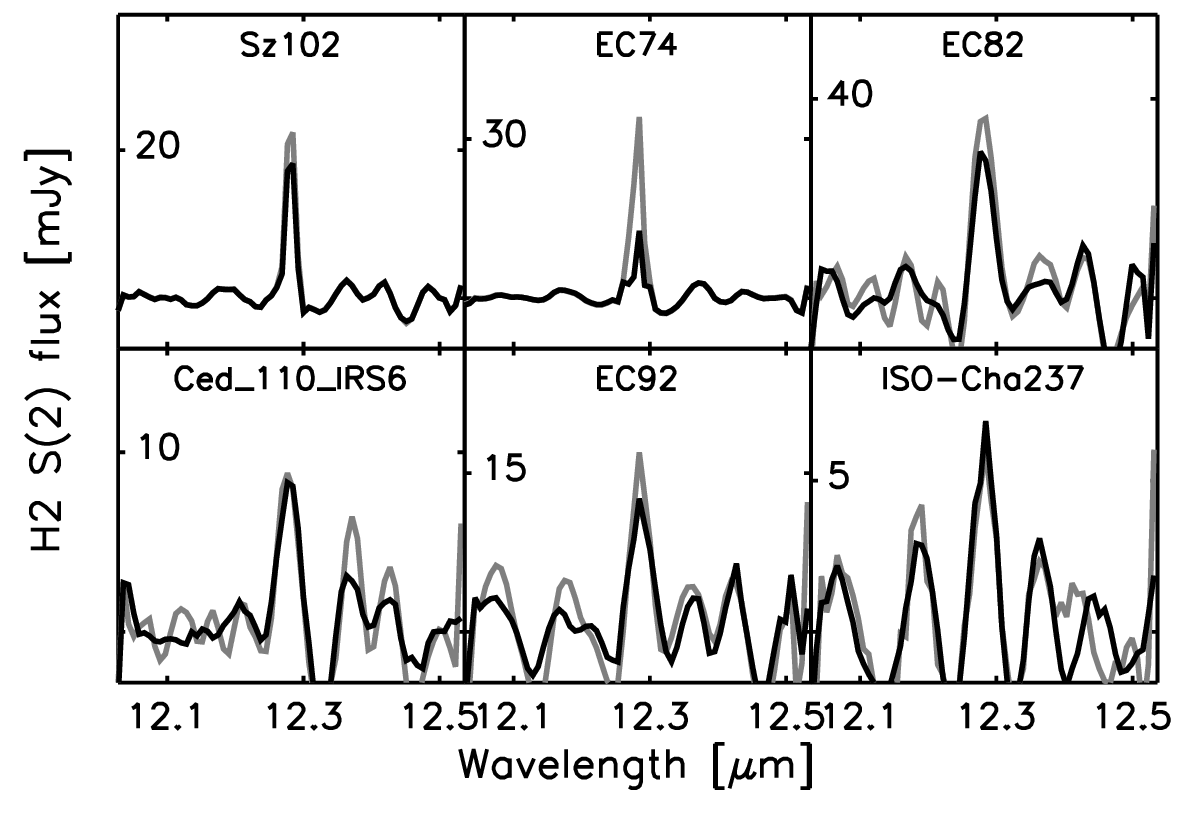}\\
\includegraphics[width=\columnwidth]{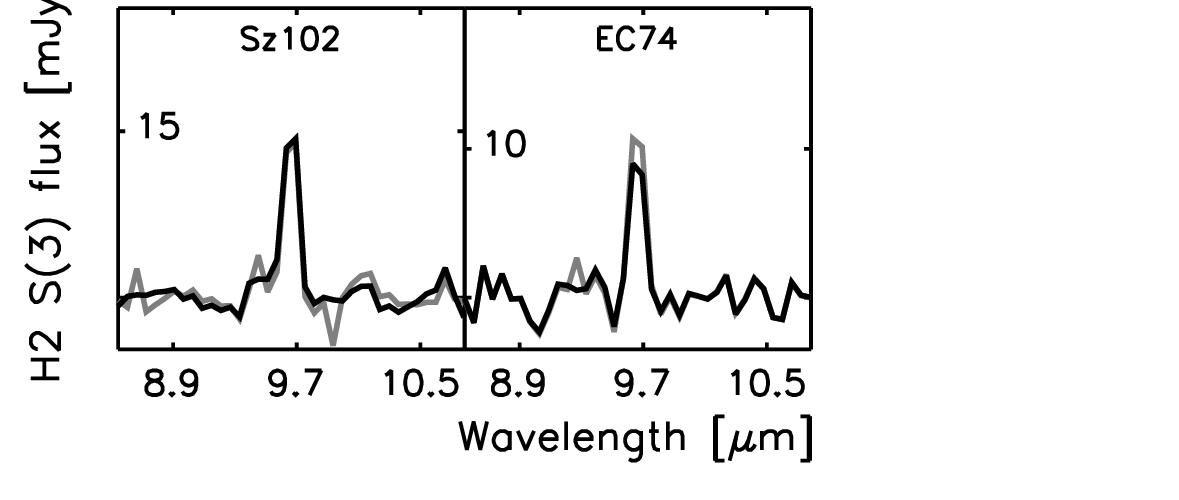}
\caption{\small\label{fig:H2_lines}%
Detection of H$_2$ lines toward the c2d sample of T~Tauri
stars with disks. Only a few of the 76 sources show clear detections.
Plotted in gray is the total observed emission (compact source +
extended component) and in black the emission after correction of the
estimated extended emission.
}
\end{center}
\end{figure}

\subsection{Molecular Hydrogen}
\label{sec:h2_analysis}

\begin{figure}
\begin{center}
\includegraphics[width=\columnwidth]{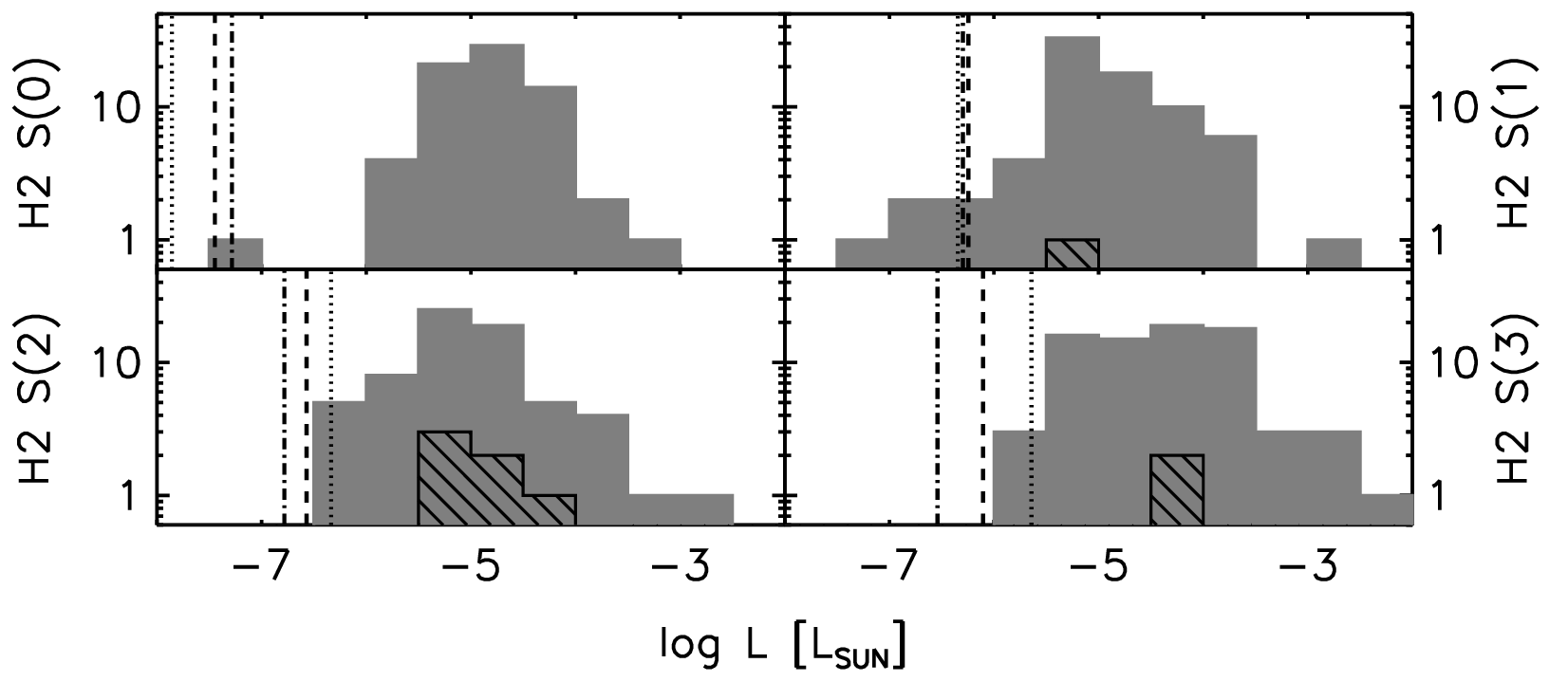}
\caption{\small\label{fig:model_lineflux}%
Observed line strengths (hatched bars) and upper limits (solid gray bars).
The vertical lines present model T~Tauri disk line strengths of
H$_2$\,S(0), S(1), S(2), and S(3) from \citet[][]{nomura07}.
The models incorporate UV and X-ray heating of the circumstellar disk for
three grain size distributions ($a_\mathrm{max}=10\mu\mathrm{m}, 1\mathrm{mm}$,
and $10\mathrm{cm}$ represented by dotted, dashed, and dashed-dotted lines
respectively). For S(0) the line strength increases with increasing
maximum grain size, while for S(2) and S(3) the line strength decreases
as the maximum grain size increases.
}
\end{center}
\end{figure}

H$_2$ emission is detected toward a small number of sources.  Figure
\ref{fig:H2_lines} shows the observed H$_2$\,0-0\,S(2) and S(3)
emission lines detected at 3$\sigma$ or more.  The total (compact and
extended line emission) observed H$_2$ emission (in gray) 
and the emission after subtraction of the extended
line emission (in black) are shown.
The H$_2$\,0-0\,S(0) and S(1) lines are seen toward some
sources such as HD\,132947 and Sz\,102 (Fig.\ \ref{fig:off-source})
but they are found to be mostly extended. Only for Sz\,102 compact
source emission of H$_2$\,0-0 S(1) is tentatively detected.
Neither S(0) nor S(1) are seen toward HD\,135344 and HD\,163296.
The 3$\sigma$ upper limits for HD\,135344 are a factor of $\sim2-3$ lower
than the tentative detections in \citet[][]{thi01}, while
for HD\,163296 the upper limits are comparable to the ISO SWS line fluxes.
The observed line fluxes and upper limits of the H$_2$\,0-0\,S(0),
S(1), S(2), and S(3) emission lines are listed in columns $4-7$
of Table\,\ref{disktab:linefluxes}. 
Figure \ref{fig:model_lineflux} shows the distribution of observed
line strengths (hatched bars) and upper limits (solid gray bars)
of H$_2$\,0-0\,S(0), S(1), S(2), and S(3).

In the simplest analysis, the H$_2$ excitation is assumed to be in
local thermal equilibrium (LTE) \citep[e.g.,][]{thi01} with an
ortho-to-para ratio determined by the kinetic temperature of the gas
\citep[following][]{sternberg99}.
For gas temperatures 100, 150, and $\geq$200\,K, the ortho-to-para
ratios are 1.6, 2.5, and 3, respectively.
Assuming optically thin emission, the integrated flux of a rotational
line $J_u \rightarrow J_l$ for a given temperature $T_\mathrm{ex}$ is

\begin{equation}
   \label{eq:F}
   F_{ul} = \frac{hc}{4\pi\lambda}
            N(\mathrm{H}_2) A_{ul} x_u \Omega
            \ \mathrm{erg\,s^{-1}\,cm^{-2}},
\end{equation}
where $\lambda$ is the wavelength of the transition, $N(\mathrm{H}_2)$ the total
column density, $A_{ul}$ the spontaneous transition probability,
and $\Omega$ the source size. For high enough densities
($n\gtrsim 10^3\mathrm{cm}^{-3}$), the population $x_u$ follows
the Boltzmann law
\begin{equation}
   \label{eq:x}
   x_u = \frac{g_\mathrm{N}(2J_u+1) e^{-E_J/kT_{\mathrm{ex}}}}
              {Q(T_{\mathrm{ex}})}
\end{equation}
where $E_J$ is the energy of the upper level, $g_\mathrm{N}$ is the nuclear
statistical weight (1 for para and 3 for ortho H$_2$), and
$Q(T_{\mathrm{ex}})$ the partition function for the given
excitation temperature $T_\mathrm{ex}$.

Using the above equations, excitation temperatures, column densities
and H$_2$ gas masses can be derived from the observed line fluxes and
upper limits.  If either S(0) or S(1) are detected an upper or lower
limit on the temperature of the warm gas is derived, but if neither
are detected a temperature of 100\,K is assumed for the warm gas.  If
two or more higher excitation lines (S(2) and higher) are detected a
temperature for the hot component is derived, while if no or only one
of the higher excitation lines is detected a temperature of 1000~K is
assumed.  For Sz\,102 and EC74 temperatures of
$T_\mathrm{hot}\sim700-800$\,K could be found for the hot component.
For all other sources no temperatures could be derived for either
component.

The column density averaged over the IRS aperture can be derived from
the above equations, given the distance to the source.
For all sources in our sample the emitting source size in the
disk is smaller than the IRS aperture (Sec. \ref{sec:optimal-extraction})
and since this is unknown a typical emitting disk region is assumed.
For the warm component a source with a radius $r=100$\,AU is assumed
and for the hot component a source with a radius $r=2$\,AU.
The derived or assumed temperature plus the (upper level) column
densities give a total column density,
which in turn gives the total H$_2$ gas mass in Jovian masses,
$M = { \pi r^2 \times N \times 2 m_\mathrm{H}} / M_\mathrm{J}
$ with $m_\mathrm{H}=1.674\cdot 10^{-24}$\,gr and
$M_\mathrm{J}=1.9\cdot 10^{30}$\,gr.

\begin{figure}[b]
\begin{center}
\includegraphics[width=\columnwidth]{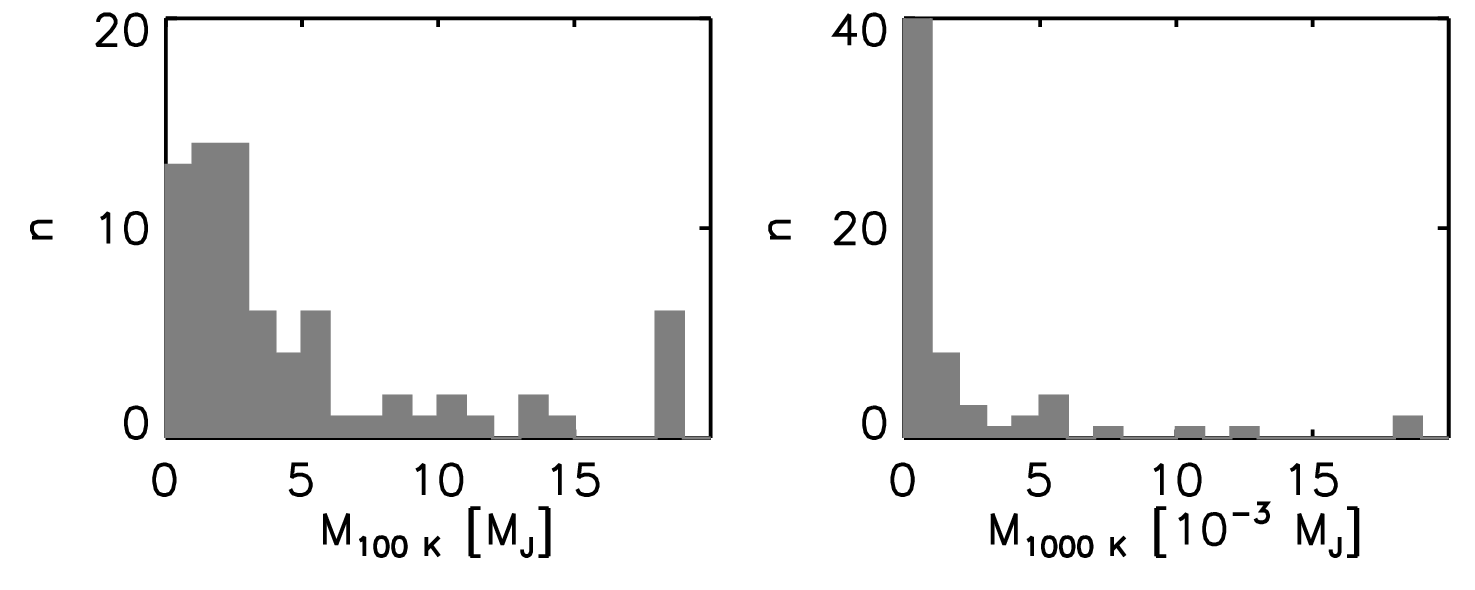}
\caption{\small\label{fig:h2_mass}%
Distribution of H$_2$ mass upper limits derived for 100\, and 1000\,K gas.
}
\end{center}
\end{figure}

The derived H$_2$ parameters for both the warm and hot gas component
are listed in Table \ref{disktab:gasparam}. Figure \ref{fig:h2_mass} shows
the distribution of the derived H$_2$ masses for the assumed temperatures
of 100\,K and 1000\,K, respectively.

\begin{figure*}
\begin{center}
\includegraphics[width=0.9\textwidth]{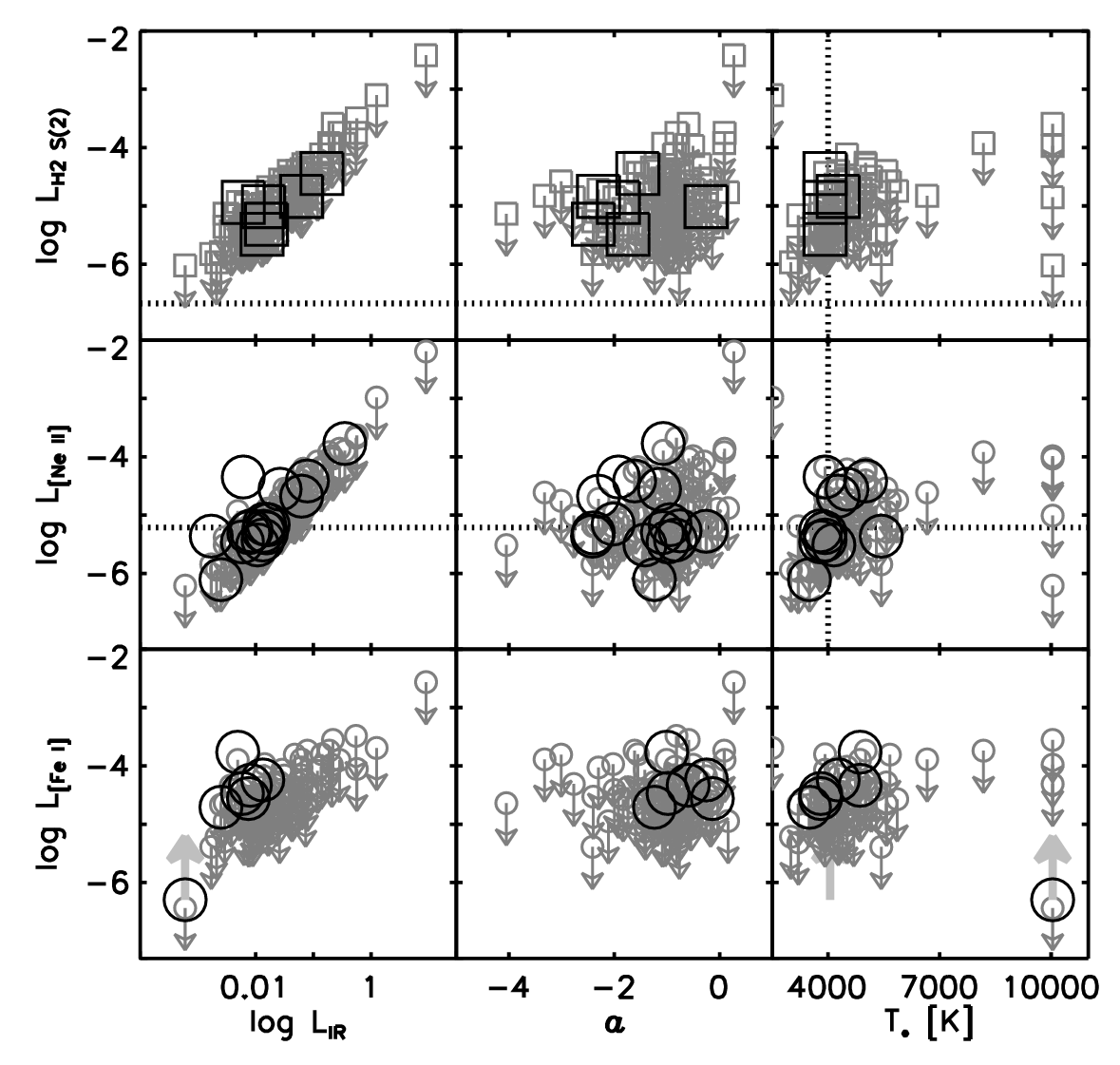}
\caption{\small\label{fig:line_relations}%
Comparison of the observed H$_2$\,S(2), [Ne\,II], and [Fe\,I] line strengts
in solar luminosities with the mid-IR luminosity, the $12.8-26\mu$m spectral
index, and the effective temperature of the sources. Plotted with the large
black symbols are the line detections, while the small gray symbols and
arrows are all non-detections.
The horizontal dotted lines in the H$_2$ and [Ne\,II] panels show the predicted
line strength by \citet[][]{nomura07} and \citet[][]{glassgold07} who use an
effective temperature of 4000\,K in their models. 4000\, is indiated with
vertical dotted lines.
}
\end{center}
\end{figure*}

\strut\\
\subsection{Correlations}
\label{sec:discussion-correlations}
Figure \ref{fig:line_relations} shows the observed line strengths and
upper limits of H$_2$\,0-0\,S(2), [Ne\,II], and [Fe\,I] as functions
of the mid-IR luminosities (in Solar luminosities), the $12.8-26\mu$m 
spectral index, and the effective temperature. 
The mid-IR luminosity is integrated using the
12.8 and $15.5\mu$m continuum points derived in the [Ne\,II]
($12.8\mu$m) and [Ne\,III] ($15.5\mu$m) spectral line fits. The
effective temperature is derived from the stellar type
\citep[][]{gray94,baraffe96}.  In all panels the sources with
H$_2$\,0-0\,S(2), [Ne\,II], or [Fe\,I] detections are plotted with
large symbols whereas the sources without line
detections are plotted with small gray symbols.

The line strength plots as function of the mid-IR luminosity clearly
show the detection limits of the \textit{Spitzer} IRS instrument with
an obvious increase of the upper limits with increasing continuum.
Therfore no definite conclusions can be drawn about the apparent
correlations, and the detection rate could be increased significantly
with higher $S/N$ for strong continuum sources. However the plots do
show that there are no low luminosity sources with strong
H$_2$\,0-0\,S(2) or [Ne\,II] emission.  No correlation with the
$12.8-26\mu$m spectral index is observed.  The detections are
distributed over a range of spectral indices, illustrating that the
line detections are not limited to a single type of disk source (e.g.,
flat or flaring).

The correlation with the effective temperature shows a differentiation
between [Ne\,II] and [Fe\,I] compared with H$_2$\,0-0\,S(2). The
[Ne\,II] and[Fe\,I] line strengths show a similar correlation with
effective temperature as with the mid-IR luminosity, but with more
scatter. Also, upper limits are seen below the correlation line for
sources with detections. There are a few detected sources deviating
from the observed trend, such as the cold disk source T\,Cha
\citep[see][]{brown07} detected in [Ne\,II] and the Herbig Ae star
HD\,132947 detected in [Fe\,I]. T\,Cha is located at a distance of
66\,pc, much closer than the majority of sources in our sample.
Sources with a similar [Ne\,II] line strength as T\,Cha but at the
distances of the nearest star-forming clouds would go undetected at
the sensitivity limits of the current sample.  For HD\,132947 the
distance is unknown and the assumed distance of 60\,pc is the lower
limit from Tycho.  The H$_2$\,0-0\,S(2) line, although detected for
only a small number of sources, differs from [Ne\,II] and [Fe\,I] in
that all sources are concentrated around a single effective
temperature. This may be real, but it could also be the result of a
S/N selection. More sensitive observations will be required to draw 
firm conclusions.


\begin{figure*}
\begin{center}
\includegraphics[width=0.4\textwidth]{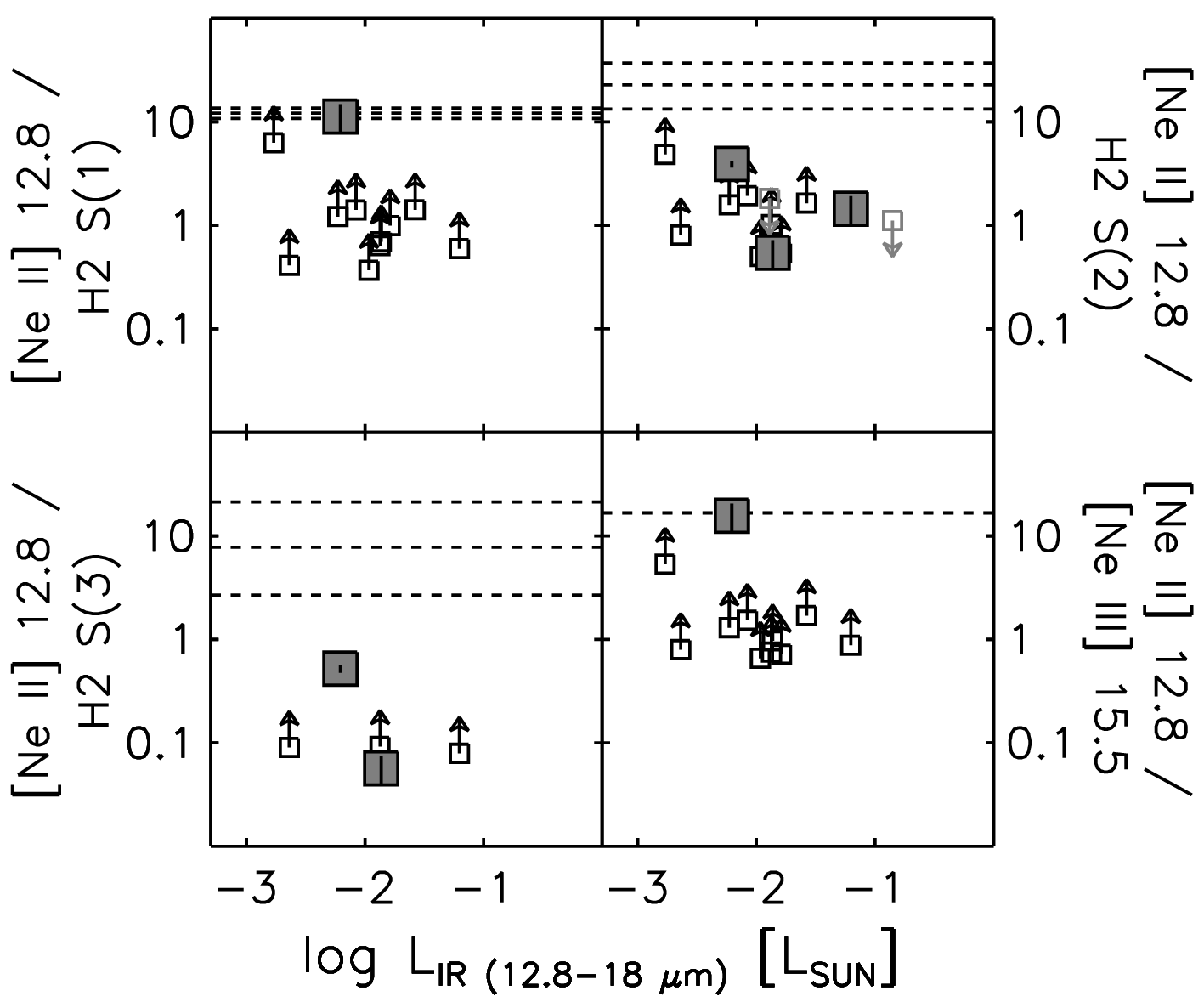}~~~~~~~~
\includegraphics[width=0.4\textwidth]{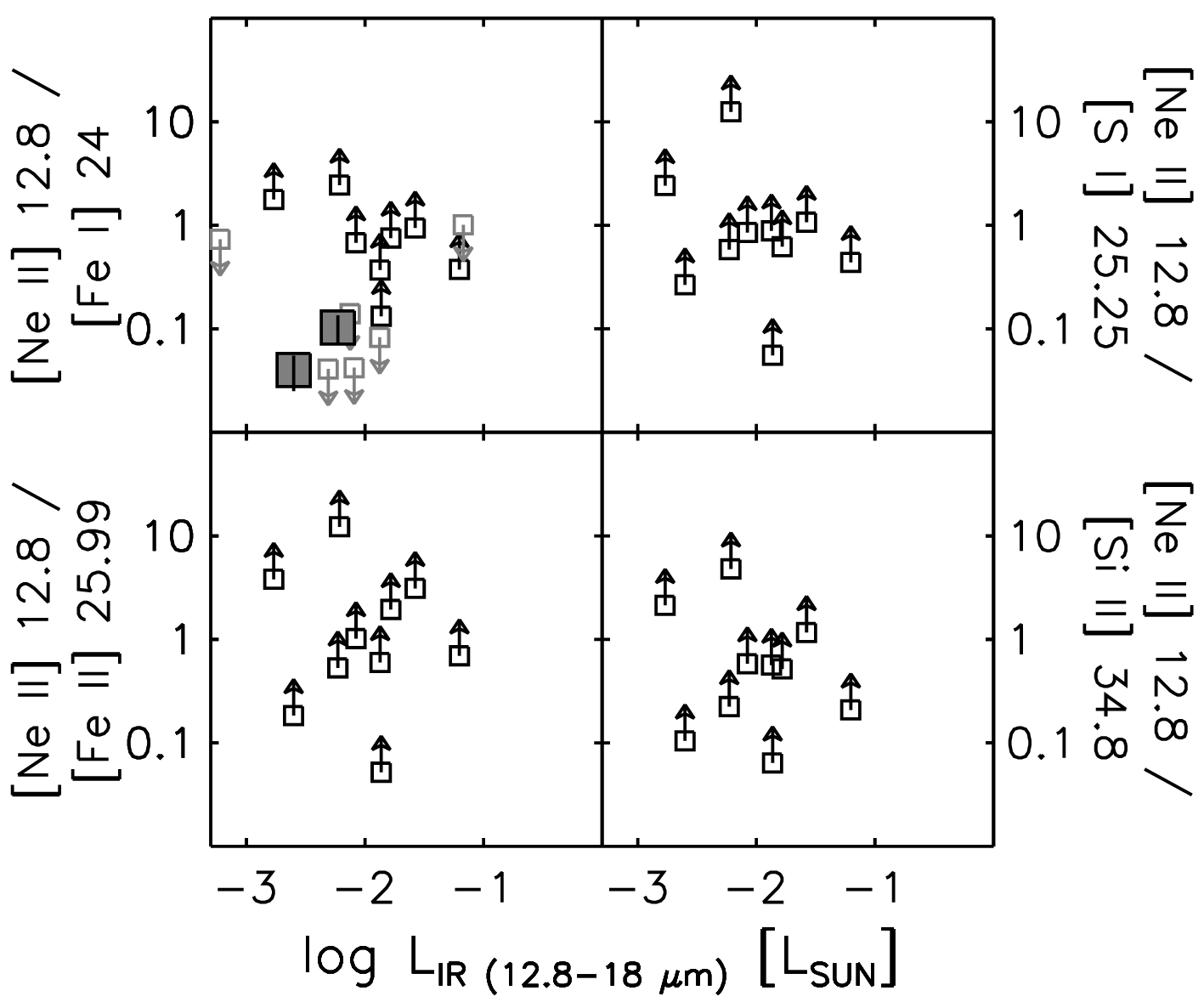}\\~\\~\\
\includegraphics[width=0.4\textwidth]{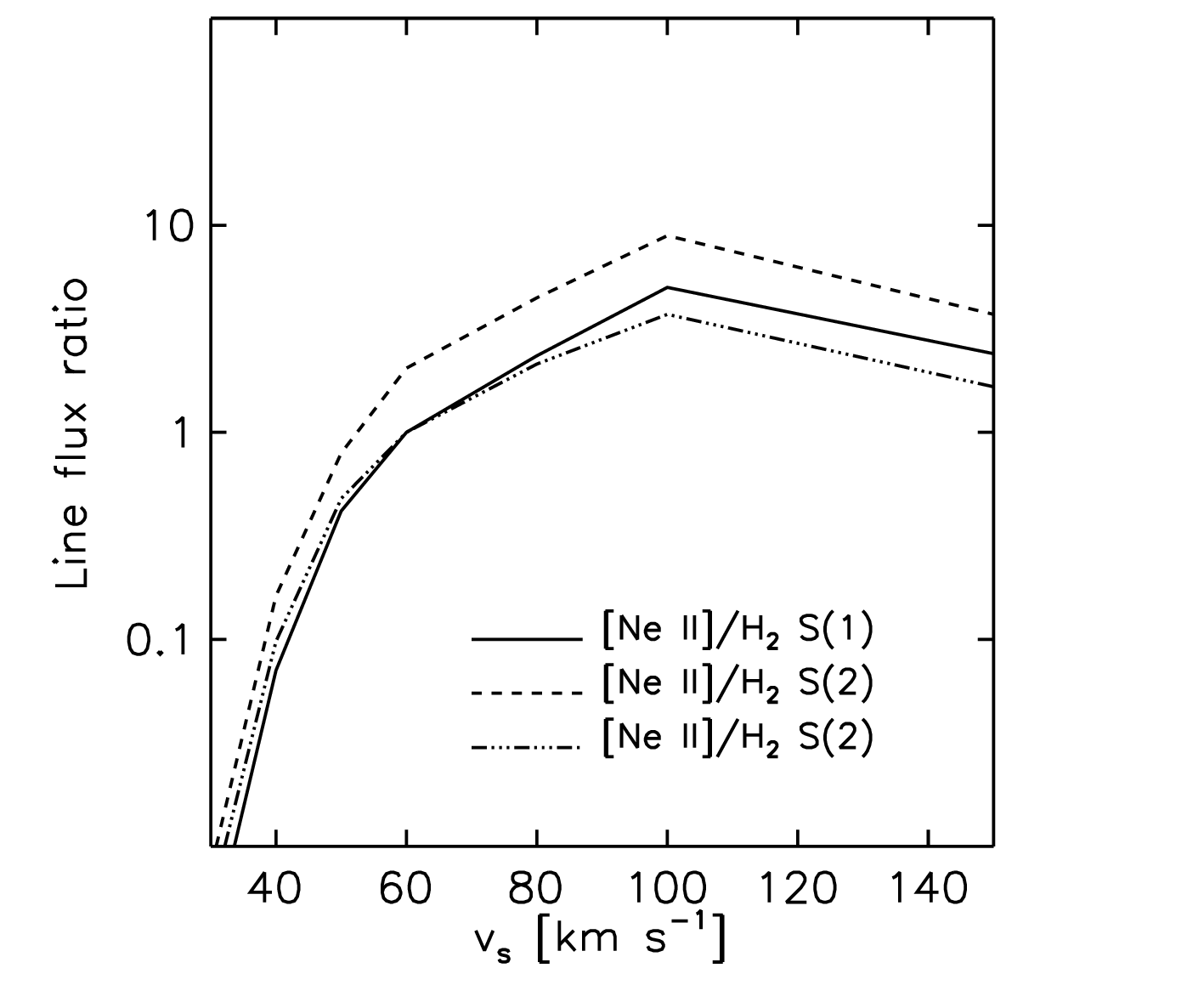}~~~~~~~~
\includegraphics[width=0.4\textwidth]{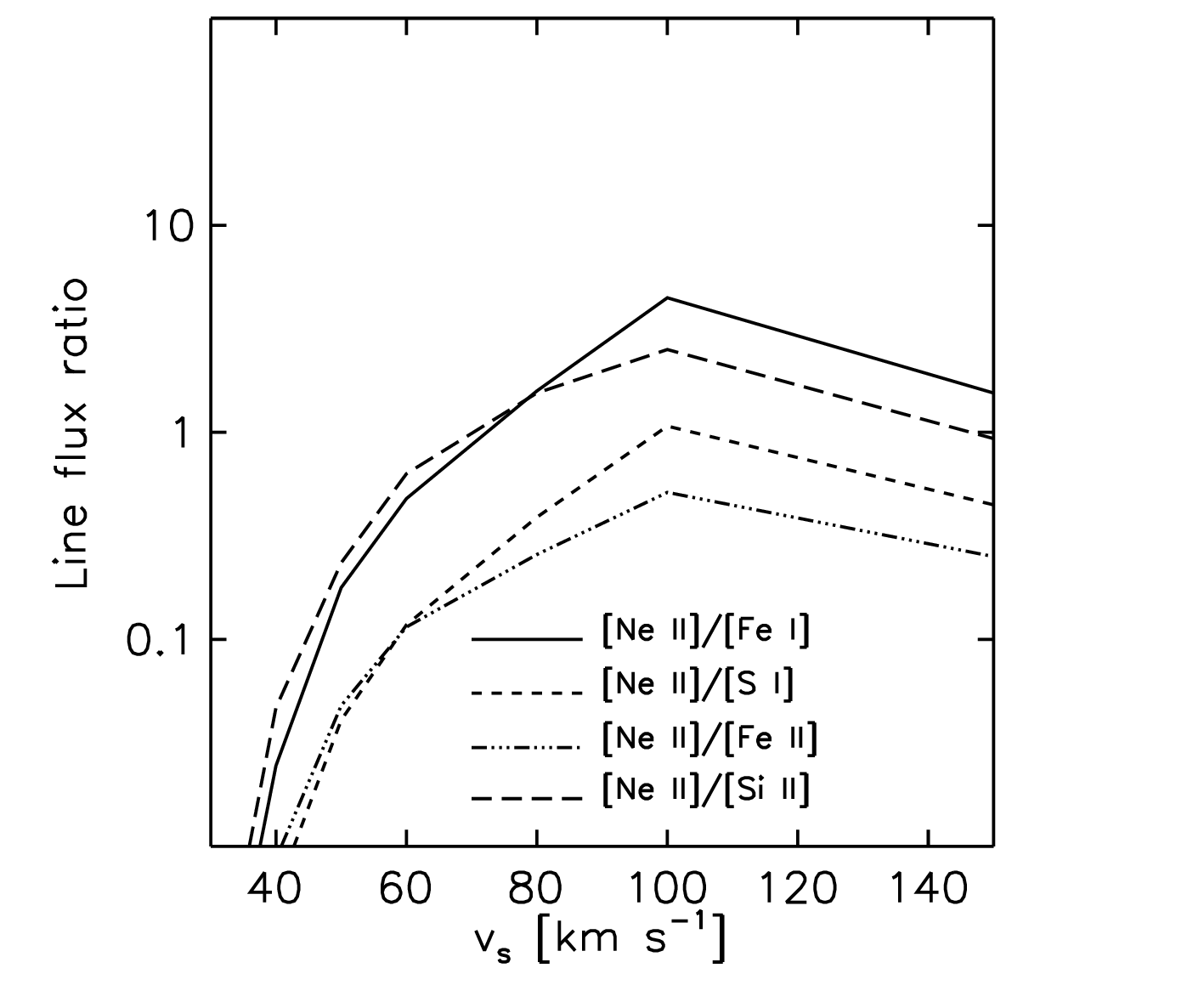}
\caption{\small\label{fig:line_ratios_neII}%
Line ratios of [Ne\,II] w.r.t. other species as functions of mid-IR
luminosity. Large filled symbols are used for sources with both
[Ne\,II] and the second line detected. Small open symbols with
arrows indicate lower and upper limits of the line ratios.
The dashed lines in the left plot show the line ratios for the
predictions from the \citet[][]{glassgold07,nomura07} models.
The lower two plots show the ratios from shock models by
\citet[][]{hollenbach89} for different shock velocities.
}
\end{center}
\end{figure*}

\begin{figure*}
\begin{center}
\includegraphics[width=0.4\textwidth]{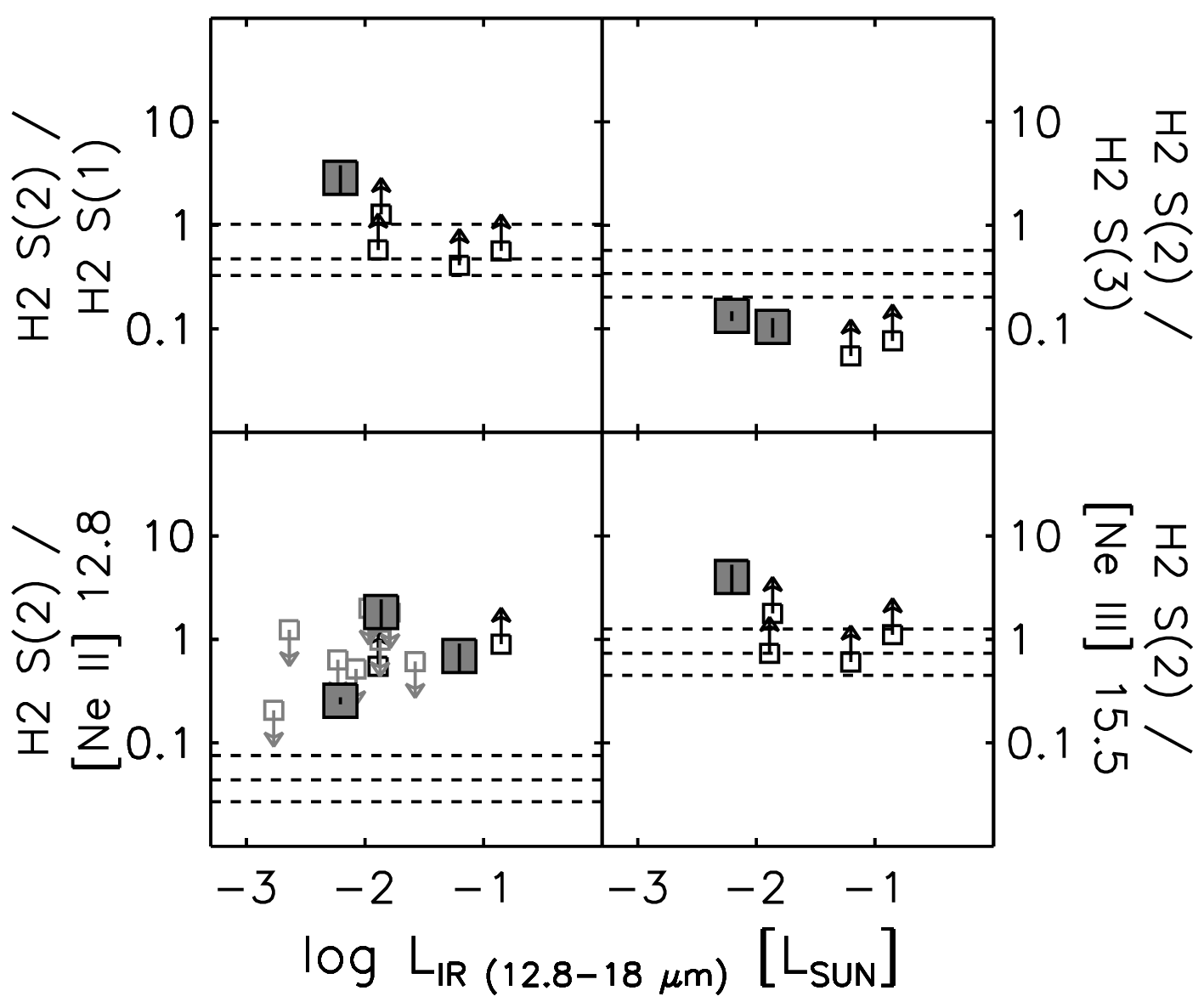}~~~~~~~~
\includegraphics[width=0.4\textwidth]{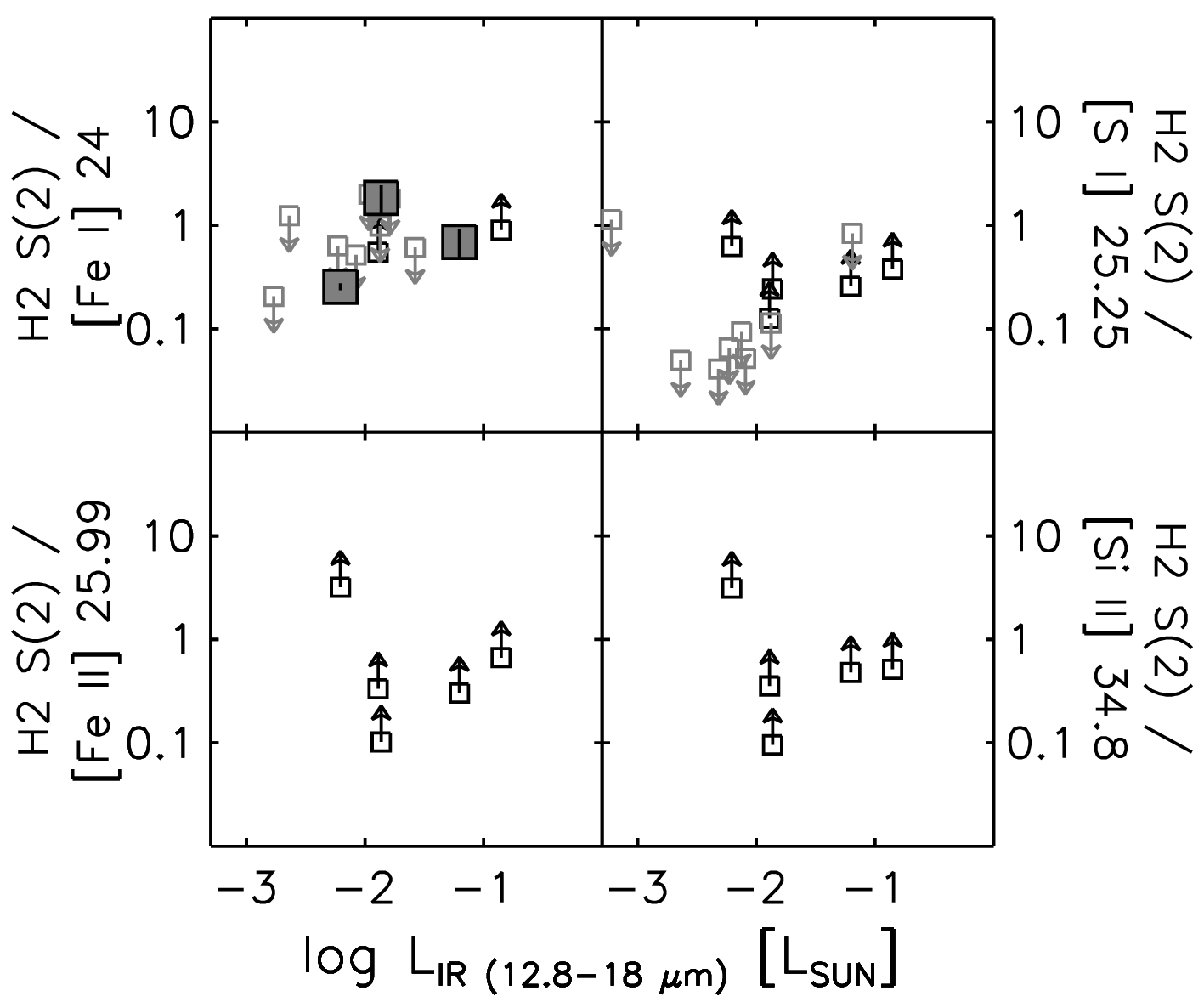}\\~\\~\\
\includegraphics[width=0.4\textwidth]{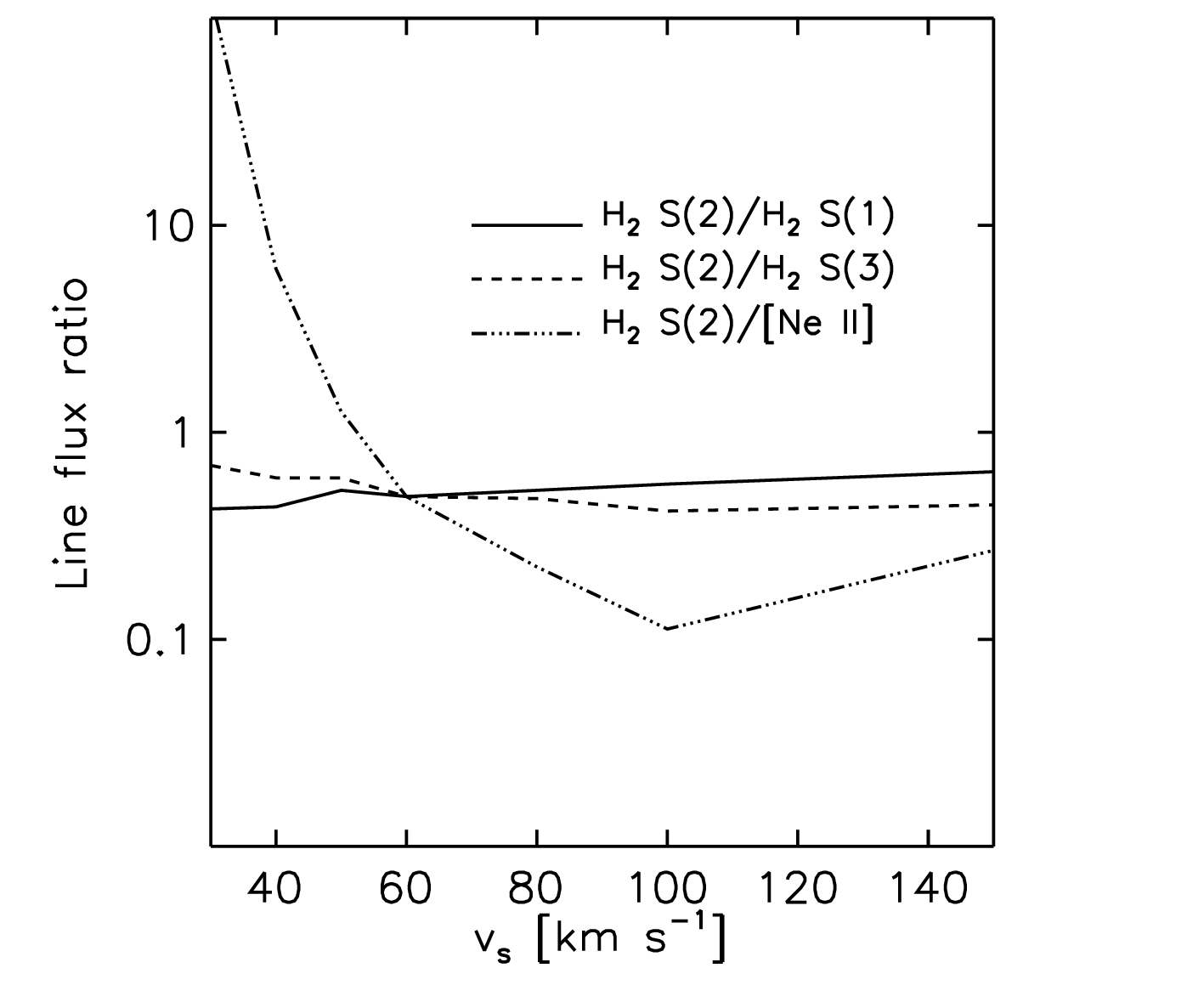}~~~~~~~~
\includegraphics[width=0.4\textwidth]{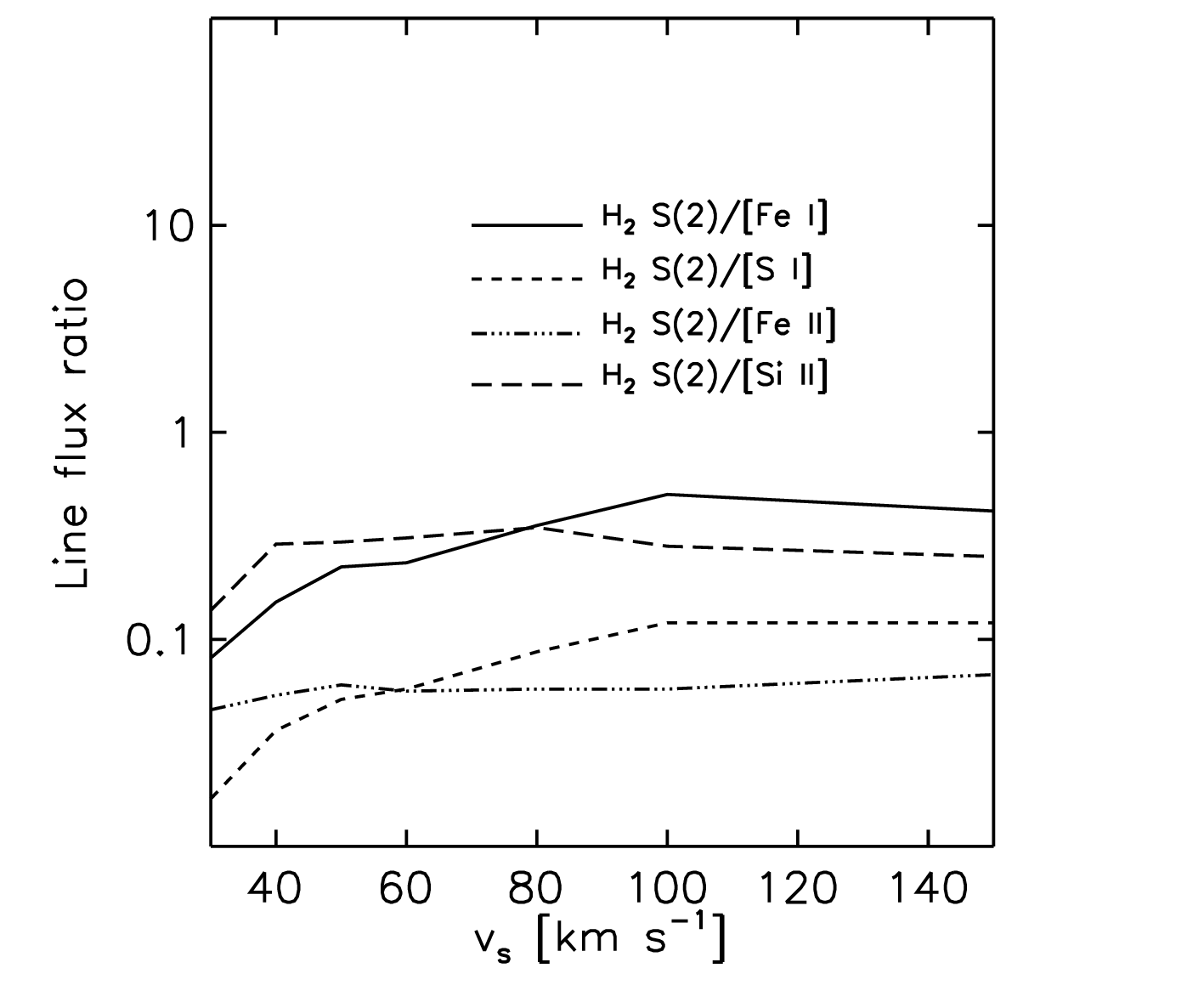}
\caption{\small\label{fig:line_ratios_h2}%
Line ratios of H$_2$ S(2) w.r.t. other species as functions of the mid-IR
luminosity.  Large filled symbols are used for sources with both
H$_2$\,S(2) and the second line detected. Small open symbols with arrows
indicate lower and upper limits of the line ratios.
The dashed lines In the left plot show the line ratios for the
predictions from the \citet[][]{glassgold07,nomura07} models.
The lower two plots show the ratios from shock models by
\citet[][]{hollenbach89} for different shock velocities.
}
\end{center}
\end{figure*}

\section{Discussion}
\label{sec:discussion}

The most significant outcome of this survey is the detection of [Ne\,II] 
emission toward 15 T~Tauri stars,
$\sim20$\,\% of the sample. [Fe\,I] is also seen toward 7 sources,
$\sim9$\,\% of the sample (see Sect. \ref{sec:neII} and
\ref{sec:other-species}).

\subsection{[Ne\,II]}
\label{sec:discussion-NeII}

Since neon cannot be ionized by photons with energies less than 21.4
eV (and Ne$^+$ 41.0\,eV), the detection of [Ne\,II] is evidence for
the presence of higher energy photons in the circumstellar
environment, specifically EUV photons or X-rays originating from
either the stellar chromosphere or (accretion) shocks. Alternatively,
high-velocity shocks can result in ionized lines.

\subsubsection{X-ray emission}

T~Tauri stars are known to be strong emitters of moderately hard X-rays.
\citet[][]{feigelson04} report
X-ray luminosities toward young stars in the Chamealeon I north cloud of
$L_\mathrm{X}=10^{28}-10^{30} \,\mathrm{erg\,s}^{-1}$
whereas \citet[][]{telleschi07} find X-ray luminosities toward young
stars in the Taurus molecular clouds of $L_\mathrm{X}=10^{28}-10^{31}
\,\mathrm{erg\,s}^{-1}$ with a clear stellar mass dependence.
A number of sources in our sample have been identified with X-ray sources
and have derived X-ray luminosities varying from
$L_\mathrm{X}\sim4\times10^{28} - 10^{31} \,\mathrm{erg\,s}^{-1}$ 
(Table \ref{disktab:selected}).

Recently \citet[][]{glassgold07} modeled the excitation of neon in an
X-ray irradiated flaring disk model according to \citet[][]{dalessio99}
and predict [Ne\,II] and [Ne\,III] line intensities. For a source at 
140\,pc (the assumed distance in \citet[][]{glassgold07}) [Ne\,II] line
fluxes of $(0.6-1)\times10^{-14}$ $\mathrm{erg\,cm^{-2}\,s^{-1}}$
and [Ne\,III] line fluxes of $(5-6.5)\times10^{-16}$
$\mathrm{erg\,cm^{-2}\,s^{-1}}$ are predicted. 
The predicted [Ne\,II] line strength of $\sim4-8\times 10^{-6}L_\sun$ falls
within the observed range of line strengths 
(see Table~\ref{fig:model_lineflux_atomic}).
The predicted [Ne\,III] line strength is at the lower end of the 
\textit{Spitzer} IRS [Ne\,III] upper limits.
Interestingly the predicted [Ne\,II] line strength agrees well at 4000\,K 
(the assumed effective temperature in the \citet[][]{glassgold07}
model) with the trend in the observed line strengths as shown
in Figure \ref{fig:line_relations}. Also the [Ne\,II]/[Ne\,III] 
line ratio for Sz\,102 agrees with the line ratios from the
\citet[][]{glassgold07} model.

The \citet[][]{glassgold07} model is based on an assumed neon abundance of
$10^{-4}$ and an X-ray luminosity and spectral temperature of
$L_\mathrm{X}=2\times10^{30}\,\mathrm{erg\,s}^{-1}$ and $kT_\mathrm{X}=1$\,keV,
appropriate for solar-mass pre-main sequence stars observed in Orion
\citep[][]{wolk05}. Lower mass, older, and accreting stars
may have a lower X-ray luminosity
\citep[e.g.][]{feigelson04,preibisch05,telleschi07} leading to lower
expected line intensities whereas higher X-ray luminosities,
higher neon abundances \citep[][]{drake05,cunha06} or the inclusion
of H atom collisions (not included in the \citet[][]{glassgold07} calculations)
may yield higher predicted line intensities. The neon line fluxes
also scale with the disk mass surface density and are therefore sensitive
to the disk geometry, e.g. flaring or non-flaring disks.

Of the sources with [Ne\,II] emission about 30\,\% are identified as
X-ray sources. The remaining sources may have escaped detection
due to incomplete or sensitivity limited X-ray searches or due
to source geometry prohibiting the detection of the X-rays.  A more
targeted deep X-ray search would be required to confirm a direct
relation between observed [Ne\,II] emission and X-ray luminosities.
Overall, variations in X-ray luminosities, age, stellar mass and
geometry appear able to cover the two to three orders of magnitude
range in the observed [Ne II] line fluxes.

\subsubsection{EUV radiation}

EUV radiation originating from the stellar chromosphere or the
accretion shock may be an additional heating component of the disk
surface and contribute to the neon excitation. EUV photons are however
quickly absorbed by atomic hydrogen in the accretion column
\citep[][]{alexander05} and for strong accretors will not reach the
disk surface.  However, for transitional objects like T~Cha the
accretion column can become optically thin to EUV photons and some EUV
radiation may escape the immediate surroundings of the star and reach
the disk surface, potentially contributing to the ionizing radiation
at the factor of 2 level. The [Ne\,II]/[Ne\,III] ratio is expected to
increase with the additional EUV contribution given the high
(41.0\,eV) ionization potential of Ne$^+$. 
For Sz\,102 the tentative [Ne\,III] detection results in a 
[Ne\,II]/[Ne\,III] ratio consistent with X-ray excitation. For all other
sources the [Ne\,III] upper limits do put any constraints on this 
(see Fig. \ref{fig:line_ratios_neII}), however.

\subsubsection{Disk shocks}
\label{sec:disk-shocks}

The presence of strong [Ne\,II] could also indicate a possible origin
of the observed [Ne\,II] emission induced by shocks in the disk.
\citet[][]{hartmann89} describe shocks resulting from the stellar wind
striking the disk surface at an oblique angle. For typical wind
velocities of 200 $\mathrm{km\,s^{-1}}$ shock velocities along the
disk surface are estimated to be $\sim$30-40
$\mathrm{km\,s^{-1}}$. For such shocks and a medium with a density of
$10^5-10^6$ $\mathrm{cm}^{-3}$ \citet[][]{hollenbach89} predict
[Ne\,II] line strengths of $\sim10^{-6}-10^{-4}$
$\mathrm{erg\,cm^{-2}\,s^{-1}\,sr^{-1}}$.  For a 100\,AU disk at
100\,pc, the upper end of this range implies a [Ne\,II] line flux of
approximately $10^{-14}$ $\mathrm{erg\,cm^{-2}\,s^{-1}}$, of the same
order as observed.  However, \citet[][]{hollenbach89} also predict the
H$_2$\,0-0\,S(1), S(2), and S(3) lines and the [Fe\,I]\,$24\mu$m,
[S\,I]\,$25.25\mu$m, [Fe\,II]\,$26\mu$m, and [Si\,II]\,$34.8\mu$m
spectral lines to be stronger than the [Ne\,II] line by $1-3$ orders
of magnitude for these velocities as illustrated in the lower right
plot of Fig.~\ref{fig:line_ratios_neII}.  The top plots of
Fig.~\ref{fig:line_ratios_neII} show the observed ratios and lower
limits. For almost all of the line pairs, the ratios are roughly equal
to or larger than unity, except for a small number of cases.


To account for line ratios equal to or larger than unity, higher
velocity shocks would be required. At high shock velocities
($v\gtrsim70\mathrm{km\,s^{-1}}$) and high densities
($\gtrsim10^5\mathrm{cm}^{-3}$) the J-shock models presented in
\citet[][]{hollenbach89} give [Ne\,II], [Fe\,I] 24 $\mu$m, [S\,I]
25.25 $\mu$m, [Fe\,II] 26 $\mu$m, and [Si\,II] 34.8 $\mu$m lines all
of comparable strength of $\sim$0.004
$\mathrm{erg\,cm^{-2}\,s^{-1}\,sr^{-1}}$ (see bottom right plot in
Figure \ref{fig:line_ratios_neII}).  To produce a line flux of
$\sim10^{-14}\ \mathrm{erg\,cm^{-2}\,s^{-1}}$ the shocked emission
would have to come from a region of the disk with a radius of
$\sim$10\,AU at a distance of 100\,pc.  Higher velocity shocks may
produce the observed line flux ratios and lower limits, but a possible
origin for such high velocity shocks is unclear.  Another problem lies
in the non-detection of the [S\,I] $25.25\mu$m, the [Fe\,II] $26\mu$m,
and the [Si\,II] $34.8\mu$m lines.  In particular the [S\,I] line
should be detected if high velocity shocks are the origin of the
observed line emission: the [S\,I] line is predicted to be stronger
whereas the detection limits for [Fe\,I] and [S\,I] are comparable in our
data (see Fig. \ref{fig:model_lineflux_atomic}).

\subsection{[Fe\,I] and [S\,I]}
\label{sec:FeI}

As mentioned in Sect.~\ref{sec:discussion-NeII} the detection of [Fe\,I]
in combination with the non-detection of other atomic lines, 
in particular [S\,I] and [Si\,II], is significant.

\citet[][]{gorti04} modeled the line emission from intermediate aged
optically thin disks around G and K stars. For disks with low gas
masses ($10^{-3}$ to $10^{-2} M_\mathrm{J}$), the
[S\,I] $25.2\mu$m, [Fe\,II] $26\mu$m, and [Si\,II] $35.4\mu$m lines
are expected to be the strongest mid-infrared emission lines.  However
as the disk mass increases a larger fraction of the sulfur turns
molecular and the [Si\,II] and [Fe\,II] emission becomes optically
thick. At the same time the [Fe\,I] lines are predicted to become
increasingly stronger and will at some point, around a $0.1M_\mathrm{J}$,
dominate over the [S\,I] emission.  Although specific calculations for
these optically thick disks are lacking, the seven sources (WX\,Cha,
C7-11, HD\,132947, LkH$\alpha$\,271, Coku\,Tau\,3, Haro\,1-17,
and XX\,Cha) which show strong [Fe\,I] emission may well have 
optically thick massive gas-rich disks.

\subsection{Molecular hydrogen}
\label{sec:discussion-H2}

The third significant result of our survey is the non-detection of
the H$_2$\,0-0\,S(0) and S(1) lines for 76 T~Tauri and Herbig Ae/Be
stars.  This puts constraints on the mass of warm ($T\sim100-200$\,K)
H$_2$ gas in the disks around these stars of typically a few $M_{\rm
J}$ as illustrated in Figure \ref{fig:h2_mass}.  Models of disk
heating by stellar UV photons show that the gas temperature in the
surface layers can be significantly higher than that of the dust down
to an optical depth for UV photons of $\sim$1. The precise
temperatures depend on the model details, in particular the presence
of PAHs, the grain size, the gas/dust ratio, and the presence of
excess UV over that of the stellar photosphere
\citep[][]{jonkheid04,kamp04,nomura05,nomura07,jonkheid06,jonkheid07}.
For interstellar-sized grains ($\sim$0.1 $\mu$m), the models readily
give surface temperatures of 100 K or more out to at least 100 AU.
Even models in which the dust grains have grown and settled to the
midplane have warm surface layers as long as some PAHs are still
present.  The total mass contained in this warm layer is however
small, $\sim1$\,\% or less of the total disk mass. For the specific
0.07 $M_\odot$ disk studied by \citet[][]{jonkheid04} the mass at
$T>100$\,K is $\sim$0.7 $M_\mathrm{J}$.  Thus, for a typical disk mass
of 0.01 M$_{\odot}$, this may be as low as 0.1 M$_{\rm J}$, below
our upper limits.

The H$_2$ line fluxes from a protoplanetary disk representative of
that around TW Hya have been modeled by \citet[][]{nomura05}.
These models include not just thermal excitation but also UV pumping of the
H$_2$ levels. 
\citet[][]{nomura07} include X-ray irradiation and the effect of grain
size distributions. For the higher excitation lines the grain size
distribution is particularly important with a distribution toward
smaller grain sizes producing higher line strengths. This result is
consistent with that of \citet[][]{jonkheid07}, who find lower
temperatures in models with grain growth.
Figure \ref{fig:model_lineflux} shows the distribution of the observed
line strengths with the predicted line strengths 
for the \citet[][]{nomura07} model included for
grain size distributions with maximum grain sizes of
$a_\mathrm{max}=10\mu\mathrm{m}, 1\mathrm{mm}$, and
$10\mathrm{cm}$. It is seen that the observed upper limits are all
consistent with this model, even if excess UV is included.  Note,
however, that pure rotational H$_2$ line fluxes are extremely
sensitive to the model details: small changes in the heating and
cooling processes, as well as the treatment of the H/H$_2$ transition
zone, can result in significant differences in gas temperatures and an
order of magnitude variation in predicted line fluxes \citep[see
discussion in][]{li02,roellig07}.  Therefore, comparison of the total
mass of warm gas between models and observations is equally relevant.

Toward six sources ($\sim$8\,\% of the sample), H$_2$\,0-0\,S(2)
and/or S(3) emission is observed, which provides evidence for the
presence of a significant hot ($T\gtrsim 500$\,K) gas component in the
disks.  Hot gas ($T\gtrsim 500$\,K) is observed toward a number of
sources through the H$_2$\,0-0\,S(2) line, most convincingly toward
Sz\,102, EC\,82, Ced\,IRS\,IRS6, EC\,74, and EC\,92, For Sz\,102 a
number of higher transition lines are also seen.  The observed
H$_2$\,0-0\,S(2) line strengths are more than a factor of 10 higher
than those predicted in \citet[][]{nomura05,nomura07} (see
Fig. \ref{fig:model_lineflux}).  Given the non-detection of the S(1)
line toward the same sources, which is predicted to have a similar
strength, this is an indication that these disks have an additional
source of emission from hot molecular hydrogen. 
The upper limits for H$_2$\,0-0\,S(2) and S(3) are higher for almost
all sources than the \citet[][]{nomura05} predicted values. An
additional hot component may therefore be present in these sources as
well below our detection limit.

None of the sources with evidence for an additional hot component show
evidence for PAH emission \citep[see][]{geers06}. Of the eight sources
with detected H$_2$\,0-0\,S(2), four show strong [Ne\,II]
emission, giving support to the idea of a common heating and
excitation mechanism through X-rays or EUV.  Of the [Ne\,II] sources
two, T~Cha and RR~Tau, show strong PAH emission.  Considering the
limitations in observing the PAH emission as described by
\citet[][]{geers06} we can at this stage draw no conclusions about the
relation between the hot H$_2$ emission, [Ne\,II] emission, and the
importance of PAHs and small grains.

An origin in a high velocity shock as discussed in Section
\ref{sec:disk-shocks} could produce the enhanced S(2) and S(3) line
strengths while keeping the line strengths of S(0) and S(1) reduced.
However, as discussed in Section \ref{sec:disk-shocks}, the main
problem with invoking shocks to explain [Ne\,II] and H$_2$ is to
accommodate both the detections of [Fe\,I] and the non-detections of
[S\,I] and [Fe\,II] $26\mu$m.  Figure \ref{fig:line_ratios_h2} shows
the observed and model line ratios with respect to H$_2$, similar
those in Fig.~\ref{fig:line_ratios_neII}.  An origin in an oblique
stellar wind shock faces the same problems as discussed for [Ne\,II]
in Section \ref{sec:disk-shocks}.

\section{Conclusions}

A survey of the mid-infrared gas phase pure rotational lines of
molecular hydrogen and a number of atomic fine structure transitions
has been carried out toward a significant sample of 76 circumstellar
disks with the \textit{Spitzer} IRS.  The principal findings include:

\begin{itemize}
\item {[Ne\,II]} is detected toward $\sim20$\,\% of the sources and
      [Ne III] tentatively in one source. 
      The [Ne\,II] detections and the [Ne\,II]/[Ne\,III] line flux
      ratio are consistent with disk heating and excitation of
      [Ne\,II] through X-rays as presented in \citet[][]{glassgold07}.
      Excitation through EUV radiation may contribute. Better
      constraints on the X-ray luminosities and [Ne\,III] fluxes
      are required to distinguish the two contributions.
\item {[Fe\,I]} is detected toward $\sim9$\,\% of the sources. No other low
      excitation atomic lines, such as [Fe II], [S\,I] and [Si\,II], are 
      detected. This suggests that these sources may possess optically thick
      disks with gas masses of at least a $0.1M_\mathrm{J}$.
\item Except for a tentative detection toward Sz\,102, no compact 
      H$_2$\,0-0\,S(0) and S(1) emission is observed toward
      any of the sources in our sample, setting limits of a few Jovian
      masses on the mass of the warm $T_\mathrm{ex}=100$\,K gas in the
      disks.  These limits are above model predictions. The H$_2$ line
      flux upper limits are also consistent with recent T~Tauri disk model
      predictions by for example \citet[][]{nomura07}. Earlier tentative
      ISO detections of H$_2$ in two Herbig Ae disks are not confirmed.
\item Hot ($T\gtrsim500$\,K) H$_2$ gas has been detected toward
      $\sim8$\,\% of the sources. Given the high upper limits for the
      rest of the sources, the fraction may be higher. The detection of 
      the hot gas suggests the presence of an additional source of hot 
      H$_2$ emission not included in the most recent disk models
      \citep[e.g.][]{nomura07}.
\item An origin of the enhanced H$_2$ emission in oblique shocks due to
      winds interacting with the disk surface is not 
      consistent with the non-detection of atomic lines, in particular
      the non-detections of [S\,I] and [Fe\,II] $26\mu$m.
\end{itemize}

The bright [Ne\,II] lines detected at 12.8 $\mu$m are excellent
targets for follow-up observations with high dispersion echelle
spectrometers on 8-10m class telescopes (TEXES, VISIR). The measured
spatial profiles and line shapes would provide exacting tests of the
X-ray mediated disk emission proposed here and could definitely rule
out any high-velocity shock mechanism.

\acknowledgments The authors would like to thank Jes J\o rgensen for
making the \textit{Spitzer} IRAC mosaics, and Hideko Nomura for
communicating her latest disk model results.  Astrochemistry in Leiden
is supported by a NWO Spinoza grant and a NOVA grant.  
Support for this work, part of the Spitzer Legacy Science Program, was 
provided by NASA through contracts 1224608, 1230779, and 1256316
issued by the Jet Propulsion Laboratory, California Institute of
Technology, under NASA contract 1407.
We thank the Lorentz Center in Leiden for hosting several meetings
that contributed to this paper.


\clearpage
\pagestyle{empty}
\clearpage
\LongTables
\begin{landscape}
\begin{deluxetable}{rlccccccccc}
\tablewidth{0pt}
\tablecaption{Source list\label{disktab:sourcelist} }
\tablehead{\colhead{\#}
           & \colhead{Source}\hfill
           & \colhead{RA}
           & \colhead{Dec}
           & \colhead{AOR}
           & \colhead{tint [s]}
           & \colhead{Class}
           & \colhead{D\tablenotemark{a}}
           & \colhead{Age}
           & \colhead{Spectral type}
           & \colhead{References\tablenotemark{b}}
           \\
           & \colhead{}\hfill
           & \colhead{}
           & \colhead{}
           & \colhead{}
           & \colhead{(SL/SH/LH)*ndith.}
           & \colhead{}
           & \colhead{[pc]}
           & \colhead{[Myr]}
           & \colhead{}
           & \colhead{}
           \\
           & \colhead{[1]}
           & \colhead{[2]}
           & \colhead{[3]}
           & \colhead{[4]}
           & \colhead{[5]}
           & \colhead{[6]}
           & \colhead{[7]}
           & \colhead{[8]}
           & \colhead{[9]}
           & \colhead{[10]}
           }
\startdata
       1 & RNO\ 15 & \hfill\RA{3}{27}{47}{.7} & \hfill\Dec{30}{12}{04}{.3}\hfill & 5633280\hfill & (14*1/31*2/60*1)*2 & TTs & 250 & -- &  & 1,2 \\
       2 & Lk\,H$\alpha$\ 270 & \hfill\RA{3}{29}{17}{.7} & \hfill\Dec{31}{22}{45}{.1}\hfill & 5634048\hfill & (14*2/31*2/60*2)*2 & TTs & 250 & 0.01 & K2.5-K7 & 3 \\
       3 & Lk\,H$\alpha$\ 271 & \hfill\RA{3}{29}{21}{.9} & \hfill\Dec{31}{15}{36}{.4}\hfill & 11827968\hfill & (14*2/121*2/60*2)*2 & TTs & 250 & -- & K3-K5 & 3 \\
       4 & Lk\,H$\alpha$\ 326 & \hfill\RA{3}{30}{44}{.0} & \hfill\Dec{30}{32}{46}{.7}\hfill & 5634304\hfill & (14*2/31*4/60*1)*2 & TTs & 250 & 0.3 & G-M0 & 3,4,5 \\
       5 & Lk\,H$\alpha$\ 327 & \hfill\RA{3}{33}{30}{.4} & \hfill\Dec{31}{10}{50}{.5}\hfill & 5634560\hfill & (14*1/6*2/14*2)*2 & TTs & 250 & 0.7-1.4 & A9-K2 & 2,3 \\
       6 & Lk\,H$\alpha$\ 330 & \hfill\RA{3}{45}{48}{.3} & \hfill\Dec{32}{24}{11}{.9}\hfill & 5634816\hfill & (14*1/31*1/60*1)*2 & CTTs & 250 & 5.7 & G3 & 2,3 \\
       7 & IRAS\ 03446+3254 & \hfill\RA{3}{47}{47}{.1} & \hfill\Dec{33}{04}{03}{.4}\hfill & 5635072\hfill & (14*1/121*2/60*1)*2 & TTs & 250 & -- &  & 1,2,6,7 \\
       8 & LkCa\ 8 & \hfill\RA{4}{24}{57}{.1} & \hfill\Dec{27}{11}{56}{.4}\hfill & 9832960\hfill & (--/31*8/60*2)*2 & CTTs & 140 & -- & M0 & 8,9 \\
\enddata
\end{deluxetable}
\clearpage
\end{landscape}

\clearpage
\clearpage
\begin{landscape}
\begin{deluxetable}{rlcrlrlrlrlrlrlrlrlrlrlrl}
\tablewidth{0pt}
\tablecaption{Observed linefluxes and $1\sigma$ uncertainties ($\mathrm{10^{-16} erg\,cm^{-2}\,s^{-1}}$)\label{disktab:linefluxes} }
\tablehead{\colhead{\#}
         & \colhead{Source}\hfill
         & \colhead{Dis.\,[pc]}
        & \multicolumn{2}{c}{H$_2$ S(0)}
        & \multicolumn{2}{c}{H$_2$ S(1)}
        & \multicolumn{2}{c}{H$_2$ S(2)}
        & \multicolumn{2}{c}{H$_2$ S(3)}
        & \multicolumn{2}{c}{[Ne II]}
        & \multicolumn{2}{c}{[Ne III]}
        & \multicolumn{2}{c}{[Fe I]}
        & \multicolumn{2}{c}{[Fe II]\,18}
        & \multicolumn{2}{c}{[Fe II]\,25}
        & \multicolumn{2}{c}{[S I]}
        & \multicolumn{2}{c}{[Si II]}
           \\
           \colhead{[1]}
         & \colhead{[2]}\hfill
         & \colhead{[3]}
        & \multicolumn{2}{c}{[4]}
        & \multicolumn{2}{c}{[5]}
        & \multicolumn{2}{c}{[6]}
        & \multicolumn{2}{c}{[7]}
        & \multicolumn{2}{c}{[8]}
        & \multicolumn{2}{c}{[9]}
        & \multicolumn{2}{c}{[10]}
        & \multicolumn{2}{c}{[11]}
        & \multicolumn{2}{c}{[12]}
        & \multicolumn{2}{c}{[13]}
        & \multicolumn{2}{c}{[14]}
           }
\startdata
1 & RNO\ 15              & $250$ &  --  & \hspace{-1em}($   48$) &  --  & \hspace{-1em}($   63$) &  --  & \hspace{-1em}($   30$) &  --  & \hspace{-1em}($  470$) &  --  & \hspace{-1em}($   85$) &  --  & \hspace{-1em}($   69$) &  --  & \hspace{-1em}($   84$) &  --  & \hspace{-1em}($   78$) &  --  & \hspace{-1em}($   44$) &  --  & \hspace{-1em}($   43$) &  --  & \hspace{-1em}($   43$) \\
2 & Lk\,H$\alpha$\ 270   & $250$ &  --  & \hspace{-1em}($   35$) &  --  & \hspace{-1em}($   42$) &  --  & \hspace{-1em}($   27$) &  --  & \hspace{-1em}($   63$) &  --  & \hspace{-1em}($   29$) &  --  & \hspace{-1em}($   21$) &  --  & \hspace{-1em}($   70$) &  --  & \hspace{-1em}($   27$) &  --  & \hspace{-1em}($   28$) &  --  & \hspace{-1em}($   35$) &  --  & \hspace{-1em}($  180$) \\
3 & Lk\,H$\alpha$\ 271   & $250$ &  --  & \hspace{-1em}($   53$) &  --  & \hspace{-1em}($   10$) &  --  & \hspace{-1em}($   12$) &  --  & \hspace{-1em}($   42$) &  --  & \hspace{-1em}($   12$) &  --  & \hspace{-1em}($    8$) &   880 & \hspace{-1em}($  130$) &  --  & \hspace{-1em}($   16$) &  --  & \hspace{-1em}($   19$) &  --  & \hspace{-1em}($   23$) &  --  & \hspace{-1em}($  150$) \\
4 & Lk\,H$\alpha$\ 326   & $250$ &  --  & \hspace{-1em}($   39$) &  --  & \hspace{-1em}($   25$) &  --  & \hspace{-1em}($   22$) &  --  & \hspace{-1em}($  110$) &  --  & \hspace{-1em}($   21$) &  --  & \hspace{-1em}($   26$) &  --  & \hspace{-1em}($   88$) &  --  & \hspace{-1em}($   24$) &  --  & \hspace{-1em}($   15$) &  --  & \hspace{-1em}($   33$) &  --  & \hspace{-1em}($  160$) \\
5 & Lk\,H$\alpha$\ 327   & $250$ &  --  & \hspace{-1em}($   63$) &  --  & \hspace{-1em}($   60$) &  --  & \hspace{-1em}($   53$) &  --  & \hspace{-1em}($  240$) &  --  & \hspace{-1em}($   64$) &  --  & \hspace{-1em}($   48$) &  --  & \hspace{-1em}($   76$) &  --  & \hspace{-1em}($   59$) &  --  & \hspace{-1em}($   34$) &  --  & \hspace{-1em}($   52$) &  --  & \hspace{-1em}($   48$) \\
\enddata
\end{deluxetable}
\clearpage
\end{landscape}

\clearpage
\pagestyle{plain}
\clearpage
\LongTables
\begin{deluxetable}{rlcccccc}
\tablewidth{0pt}
\tabletypesize{\scriptsize}
\tablecaption{\label{disktab:gasparam}Diagnostic parameters of the disk gas contents.}
\tablehead{\colhead{\#\tablenotemark{a}}
           & \colhead{Source}\hfill
           & \colhead{$T_\mathrm{warm}$\tablenotemark{b}}
           & \colhead{$N_\mathrm{warm}$\tablenotemark{b}}
           & \colhead{$M_\mathrm{warm}$\tablenotemark{b}}
           & \colhead{$T_\mathrm{hot}$\tablenotemark{b}}
           & \colhead{$N_\mathrm{hot}$}
           & \colhead{$M_\mathrm{hot}$}
         \\
           \colhead{}
           & \colhead{}
           & \colhead{}
           & \colhead{($r=100$\,AU)\tablenotemark{c}}
           & \colhead{}
           & \colhead{}
           & \colhead{($r=2$\,AU)\tablenotemark{c}}
           & \colhead{}
         \\
           \colhead{}
           & \colhead{}
           & \colhead{[K]}
           & \colhead{[$10^{22}\mathrm{cm^{-2}}$]}
           & \colhead{[$\mathrm{M_J}$]}
           & \colhead{[K]}
           & \colhead{[$10^{22}\mathrm{cm^{-2}}$]}
           & \colhead{[$10^{-3}\mathrm{M_J}$]}
         \\
           \colhead{[1]}
           & \colhead{[2]}
           & \colhead{[3]}
           & \colhead{[4]}
           & \colhead{[5]}
           & \colhead{[6]}
           & \colhead{[7]}
           & \colhead{[8]}
          }
\startdata
1 & RNO\,15              & $$ & $<  74$ & $<  9.2$ & $$ & $<      75$ & $<  3.7$ \\
2 & Lk\,H$\alpha$\,270   & $$ & $<  59$ & $<  7.3$ & $$ & $<      18$ & $<  0.9$ \\
3 & Lk\,H$\alpha$\,271   & $$ & $<  35$ & $<  4.3$ & $$ & $<      16$ & $<  0.8$ \\
4 & Lk\,H$\alpha$\,326   & $$ & $<  44$ & $<  5.5$ & $$ & $<      37$ & $<  1.9$ \\
5 & Lk\,H$\alpha$\,327   & $$ & $<  86$ & $< 10.6$ & $$ & $<      80$ & $<  4.0$ \\
6 & Lk\,H$\alpha$\,330   & $$ & $< 110$ & $< 13.0$ & $$ & $<      72$ & $<  3.5$ \\
7 & IRAS\,03446+3254     & $$ & $<  43$ & $<  5.3$ & $$ & $<      30$ & $<  1.5$ \\
8 & LkCa\,8              & $$ & $<   8$ & $<  1.0$ & $$ & $<       4$ & $<  0.2$ \\
\enddata
\tablenotetext{a}{100\,K is assumed for the warm component
                  and 1000\,K for the hot component.
                  The derived column density and mass of the
                  warm component depend strongly on the assumed
                  temperature. A temperature of 150 and 200\,K
                  reduces the column density and mass by
                  respectively a factor of $\sim$ 30 and 140.
                  For the hot component 1500\,K instead of
                  1000\,K may results in a reduction of column
                  density and mass up to a factor of 10.}
\tablenotetext{b}{For the unresolved emission a source with
                  $r=50$\,AU is assumed to obtain an estimate
                  of the column density.}
\end{deluxetable}

\vspace{5\baselineskip}
\begin{deluxetable}{rlccccccccc}
\tablewidth{0pt}
\tabletypesize{\scriptsize}
\tablecaption{Summary of data for sources with detected H$_2$, [Ne\,II], or [Fe\,I] emission\label{disktab:selected} }
\tablehead{
           \colhead{\#}
         & \colhead{Source}\hfill
         & \colhead{log\,$L$}
         & \colhead{log\,$L$}
         & \colhead{log\,$L$}
         & \colhead{log\,$L$}
         & \colhead{log\,$L$}
         & \colhead{T$_\mathrm{*}$}
         & \colhead{Age}
         & \colhead{Class}
         & \colhead{log $L_\mathrm{X}$\tablenotemark{a}}
        \\
           \colhead{}
         & \colhead{}
         & \colhead{H$_2$\,S(2)}
         & \colhead{H$_2$\,S(3)}
         & \colhead{[Ne\,II]}
         & \colhead{[Fe\,I]}
         & \colhead{IR}
         & \colhead{}
         & \colhead{}
         & \colhead{}
         & \colhead{}
        \\
           \colhead{}
         & \colhead{}
         & \colhead{[$L_\odot$]}
         & \colhead{[$L_\odot$]}
         & \colhead{[$L_\odot$]}
         & \colhead{[$L_\odot$]}
         & \colhead{[$L_\odot$]}
         & \colhead{[K]}
         & \colhead{[Myr]}
         & \colhead{}
         & \colhead{[erg\,s$^{-1}$]}
          }
\startdata
   27 &                     C7-11 & $   <-5.6$  & $       -$  & $   <-5.7$  & $    -4.3$  &    -2.1 &    4860 &     0.6 &     TTs &            -- \\ 
   21 &            Ced\ 110\ IRS6 & $    -5.3$  & $       -$  & $    -5.3$  & $   <-4.8$  &    -1.8 &      -- &      -- &       1 &  28.6$^{(1)}$ \\ 
   13 &              CoKu\ Tau\ 3 & $   <-5.6$  & $       -$  & $   <-5.4$  & $    -4.6$  &    -2.1 &    3800 &      -- &    CTTs &  30.8$^{(2)}$ \\ 
   71 &                      EC74 & $    -5.0$  & $    -4.0$  & $    -5.3$  & $   <-4.4$  &    -1.9 &      -- &      -- &       2 &  30.5$^{(3)}$ \\ 
   72 &                      EC82 & $    -4.4$  & $   <-3.3$  & $   <-4.4$  & $   <-4.0$  &    -0.9 &    3918 &      -- &     TTs &            -- \\ 
   74 &                      EC92 & $    -4.8$  & $   <-3.6$  & $    -4.7$  & $   <-4.3$  &    -1.2 &    4273 &      -- &     TTs &            -- \\ 
   37 &                HD\ 132947 & $   <-6.2$  & $   <-5.5$  & $   <-6.4$  & $    -6.3$  &    -3.2 &   10040 &      -- &   HAeBe &            -- \\ 
   61 &                Haro\ 1-17 & $   <-6.0$  & $   <-5.1$  & $    -6.1$  & $    -4.7$  &    -2.6 &    3500 &      -- &     TTs &            -- \\
\enddata
\tablenotetext{a}{X-ray luminosities taken from: 
(1) \citet{carkner98},
(2) \citet{audard06},
(3) \citet{preibisch03},
(4) \citet{feigelson04},
(5) \citet{gondoin06},
(6) \citet{alcala97},
(7) \citet{stelzer04}.
                }
\end{deluxetable}

\end{document}